\documentclass[a4paper,fleqn]{cas-sc}
\usepackage{color}
\usepackage{amssymb}
\usepackage{rotating}
\usepackage{lscape}
\usepackage{xcolor}
\usepackage{subcaption}
\usepackage{caption}
\usepackage{float}
\usepackage{graphicx}
\usepackage[ruled,vlined]{algorithm2e}
\usepackage[autostyle]{csquotes}
\usepackage{acro}
\usepackage{siunitx}
\usepackage{enumitem}
\usepackage[square,numbers]{natbib}
\usepackage{verbatim}
\usepackage{multirow}
\usepackage{amsmath}
\usepackage{framed} % Framing content
\usepackage{nomencl} % Nomenclature package
\usepackage[acronym,nomain,nonumberlist]{glossaries}
\usepackage[colorinlistoftodos,prependcaption]{todonotes}
\makenomenclature

\usepackage{multicol}

\usepackage{multicol} % Multiple columns environment

\DeclareAcronym{iot}{
  short = IoT , long  = internet of things , tag = abbrev
}
\DeclareAcronym{co2}{
  short =  CO$_{2}$ , long  = carbon dioxide , tag = abbrev
}
\DeclareAcronym{ml}{
  short =  ML , long  = machine learning , tag = abbrev
}
\DeclareAcronym{dl}{
  short =  DL , long  = deep learning , tag = abbrev
}
\DeclareAcronym{rgb}{
  short =  RGB , long  = red green blue , tag = abbrev
}
\DeclareAcronym{cnn}{
  short =  CNN , long  = convolutional neural network , tag = abbrev
}
\DeclareAcronym{mts}{
  short =  MTS , long  = multivariate time-series , tag = abbrev
}
\DeclareAcronym{knn}{
  short =  kNN , long  = k-nearest neighbor , tag = abbrev
}
\DeclareAcronym{dt}{
  short =  DT , long  = decision tree , tag = abbrev
}
\DeclareAcronym{rf}{
  short =  RF , long  = random forest , tag = abbrev
}
\DeclareAcronym{svm}{
  short =  SVM , long  = support vector machine , tag = abbrev
}
\DeclareAcronym{dnn}{
  short =  DNN , long  = deep neural networks , tag = abbrev
}
\DeclareAcronym{pca}{
  short =  PCA , long  = principal component analysis , tag = abbrev
}
\DeclareAcronym{lda}{
  short =  LDA , long  = linear discriminant analysis , tag = abbrev
}
\DeclareAcronym{ann}{
  short =  ANN , long  = artificial neural networks , tag = abbrev
}
\DeclareAcronym{lr}{
  short =  LR , long  = logistics regression , tag = abbrev
}
\DeclareAcronym{mtf}{
  short =  MTF , long  = Markov transition fields , tag = abbrev
}
\DeclareAcronym{gaf}{
  short =  GAF , long  = Gramian angular fields , tag = abbrev
}
\DeclareAcronym{mam}{
  short =  MAM , long  = moving average mapping , tag = abbrev
}
\DeclareAcronym{dmam}{
  short =  DMAM , long  = double moving average mapping , tag = abbrev
}
\DeclareAcronym{gadf}{
  short =  GADF , long  = Gramian angular difference fields , tag = abbrev
}
\DeclareAcronym{gasf}{
  short =  GASF , long  = Gramian angular summation fields , tag = abbrev
}
\DeclareAcronym{eeg}{
  short =  EEG , long  = electroencephalogram , tag = abbrev
}
\DeclareAcronym{ecg}{
  short =  ECG , long  = electrocardiogram , tag = abbrev
}
\DeclareAcronym{rqa}{
  short =  RQA , long  = recurrence quantification analysis , tag = abbrev
}
\DeclareAcronym{de}{
  short =  DE , long  = dispersion entropy , tag = abbrev
}
\DeclareAcronym{rem}{
  short =  REM , long  = rapid eye movement , tag = abbrev
}
\DeclareAcronym{nrem}{
  short =  NREM , long  = non-rapid eye movement , tag = abbrev
}
\DeclareAcronym{tp}{
  short =  TP , long  = true positive , tag = abbrev
}
\DeclareAcronym{fp}{
  short =  FP , long  = false positive , tag = abbrev
}
\DeclareAcronym{tn}{
  short =  TN , long  = true negative , tag = abbrev
}
\DeclareAcronym{fn}{
  short =  FN , long  = false negative , tag = abbrev
}
\DeclareAcronym{rpm}{
  short =  RPM , long  = relative position matrix , tag = abbrev
}
\DeclareAcronym{sd}{
  short =  SD , long  = standard deviation , tag = abbrev
}
\DeclareAcronym{gps}{
  short = GPS , long  = global positioning system, tag = abbrev
}
\DeclareAcronym{1d}{
  short = 1D , long  = one dimensional , tag = abbrev
}
\DeclareAcronym{2d}{
  short = 2D , long  = two dimensional , tag = abbrev
}
\DeclareAcronym{odd}{
  short =  ODD , long  = occupancy detection dataset , tag = abbrev
}
\DeclareAcronym{eco}{
  short =  ECO , long  = electricity consumption and occupancy , tag = abbrev
}
\DeclareAcronym{niom}{
  short =  NIOM , long  = non-intrusive occupancy monitoring , tag = abbrev
}
\DeclareAcronym{refit}{
  short =  REFIT , long  = retrofit decision support tools for UK homes , tag = abbrev
}
\DeclareAcronym{uci}{
  short =  UCI , long  = university of california irvine , tag = abbrev
}
\DeclareAcronym{dred}{
  short =  DRED , long  = Dutch residential energy dataset , tag = abbrev
}
\DeclareAcronym{rae}{
  short =  RAE , long  = rainforest automation energy , tag = abbrev
}
\DeclareAcronym{srd}{
  short =  SRD , long  = student room dataset , tag = abbrev
}
\DeclareAcronym{lrd}{
  short =  LRD , long  = living room dataset , tag = abbrev
}
\DeclareAcronym{qud}{
  short =  QUD , long  = Qatar university dataset , tag = abbrev
}
\DeclareAcronym{gpu}{
  short = GPU , long  = graphics processing unit , tag = abbrev
}
\DeclareAcronym{tl}{
  short = TL , long  = transfer learning , tag = abbrev
  }
\DeclareAcronym{lstm}{
  short = LSTM , long  = long short-term memory , tag = abbrev 
}
\DeclareAcronym{ami}{
  short = AMI , long  = advanced metering infrastructure , tag = abbrev 
}

\DeclareAcronym{drs}{
  short = DRS , long  = doppler radar sensors , tag = abbrev 
}
\DeclareAcronym{ita}{
  short = AMI , long  = infrared thermal
array sensors , tag = abbrev 
}

\DeclareAcronym{t}{
  short = T , long  = temperature, tag = nomencl
}
\DeclareAcronym{l}{
  short = L , long  = light-level, tag = nomencl
}
\DeclareAcronym{h}{
  short = H , long  = humidity, tag = nomencl
}
\DeclareAcronym{hr}{
  short = HR , long  = humidity ratio, tag = nomencl
}
\DeclareAcronym{a}{
  short = A , long  = altitude, tag = nomencl
}
\DeclareAcronym{ap}{
  short = AP , long  = atmospheric pressure, tag = nomencl
}
\DeclareAcronym{ws}{
  short = WS , long  = wind speed, tag = nomencl
}
\DeclareAcronym{m}{
  short = M , long  = motion, tag = nomencl
}
\DeclareAcronym{c}{
  short = C , long  = chair, tag = nomencl
}
\DeclareAcronym{d}{
  short = D , long  = door, tag = nomencl
}
\DeclareAcronym{v}{
  short = V , long  = voltage, tag = nomencl
}
\DeclareAcronym{i}{
  short = I , long  = current, tag = nomencl
}
\DeclareAcronym{np}{
  short = Np , long  = normalized power, tag = nomencl
}
\DeclareAcronym{s}{
  short = S , long  = apparent power, tag = nomencl
}
\DeclareAcronym{p}{
  short = P , long  = active power, tag = nomencl
}
\DeclareAcronym{q}{
  short = Q , long  = reactive power, tag = nomencl
}
\DeclareAcronym{e}{
  short = E , long  = energy, tag = nomencl
}
\DeclareAcronym{phase}{
  short = $\phi$ , long  = phase angle, tag = nomencl
}
\DeclareAcronym{f}{
  short = f , long  = frequency, tag = nomencl
}
\DeclareAcronym{pf}{
  short = pf , long  = power factor, tag = nomencl
}
\DeclareAcronym{ec}{
  short = EC , long  = energy cost, tag = nomencl
}
\DeclareAcronym{o}{
  short = O , long  = occupancy , tag = nomencl
}

\usepackage{svg}

\newcommand{\orcid}[1]{\href{https://orcid.org/#1}{\includesvg[width=8pt]{orcid}}}

\usepackage{xr-hyper}
\usepackage{hyperref} %<--- Load after everything else

\begin{document}
\let\WriteBookmarks\relax
\def\floatpagepagefraction{1}
\def\textpagefraction{.001}

% floatsetup[table]{font=rm}

\shorttitle{ } 

\shortauthors{Y. Himeur et al.}  

\title [mode = title] {Edge AI for Internet of Energy: Challenges and Perspectives}

%\author[1]{Yassine Himeur}[
%    orcid=0000-0001-8904-5587]   \cormark[1] 
%\ead{yhimeur@ud.ac.ae}

%\author[2]{Aya Sayed}[
%    orcid=0000-0003-0520-2149]
%\ead{as1516645@qu.edu.qa}

%\author[3]{Abdullah Alsalemi}[
%    orcid=0000-0001-7574-4766]
%\ead{abdullah.alsalemi@my365.dmu.ac.uk }
  
%\author[2]{Faycal Bensaali}[
%    orcid=0000-0002-9273-4735]
%\ead{f.bensaali@qu.edu.qa}

%\author[3,4]{Abbes Amira}[
%    orcid=0000-0003-1652-0492]
%\ead{aamira@sharjah.ac.ae }

%\address[1]{College of Engineering and Information Technology, University of Dubai, Dubai, UAE}
%\address[2]{Department of Electrical Engineering, Qatar University, Doha, Qatar}
%\address[3]{Institute of Artificial Intelligence, De Montfort University, Leicester, United Kingdom}
%\address[4]{Department of Computer Science, University of Sharjah, UAE}


\begin{abstract}
%As advances in the Internet of Energy (IoE) and Artificial Intelligence (AI) permeate the energy sector, along with the ubiquitous use of smartphone devices, a cutting-edge technology known as edge AI is gaining traction. Energy systems, households, and public buildings generate a tremendous amount of data every second, necessitating high computational capabilities to ensure real-time monitoring and actuation for energy-saving endeavors. A robust IoE system should support a variety of applications such as energy consumption tracking and prediction, appliance control, home automation, demand elasticity, non-technical loss (NTL) detection, and decentralized energy production. In this scenario, IoE devices need to communicate amongst themselves, making data transmission through cloudlet services potentially inefficient. Given the surge in end-user device usage, there is a growing need to relocate intelligence and computing tasks closer to these devices, thereby better serving user needs. In response to this, the present paper, to the authors' best knowledge, offers the first comprehensive survey of edge AI frameworks for IoE. A well-defined taxonomy serves as a basis for a thorough overview, commencing with the advantages of edge AI and its applications. Techniques for rapid inference and training at the edge are extensively discussed. This is followed by a critical analysis that sheds light on both resolved and unresolved issues, paving the way for future prospects aimed at enhancing edge AI solutions for IoE applications.
The digital landscape of the Internet of Energy (IoE) is on the brink of a revolutionary transformation with the integration of edge Artificial Intelligence (AI). This comprehensive review elucidates the promise and potential that edge AI holds for reshaping the IoE ecosystem. Commencing with a meticulously curated research methodology, the article delves into the myriad of edge AI techniques specifically tailored for IoE. The myriad benefits, spanning from reduced latency and real-time analytics to the pivotal aspects of information security, scalability, and cost-efficiency, underscore the indispensability of edge AI in modern IoE frameworks. As the narrative progresses, readers are acquainted with pragmatic applications and techniques, highlighting on-device computation, secure private inference methods, and the avant-garde paradigms of AI training on the edge. A critical analysis follows, offering a deep dive into the present challenges including security concerns, computational hurdles, and standardization issues. However, as the horizon of technology ever expands, the review culminates in a forward-looking perspective, envisaging the future symbiosis of 5G networks, federated edge AI, deep reinforcement learning, and more, painting a vibrant panorama of what the future beholds. For anyone vested in the domains of IoE and AI, this review offers both a foundation and a visionary lens, bridging the present realities with future possibilities.
\end{abstract}

% \begin{highlights}
% \end{highlights}

\begin{keywords}

Edge AI \\ 
Internet of energy (IoE) \\ 
Energy efficiency in buildings \\ 
Federated learning \\
Blockchain \\
Large language models (LLMs)

%\textbf{Word count is: \quickwordcount{main}}

\end{keywords}

\maketitle

%\clearpage

%{
%\scriptsize
%\tableofcontents
%}

%\clearpage

\section{Introduction}\label{sec1}
One critical issue under contemporary investigation relates to reducing carbon dioxide emissions through the optimization of power consumption, particularly in the building energy sector \cite{copiaco2023innovative}. In response, research institutions and industrial firms are making concerted efforts to develop innovative technology-based solutions aimed at curbing emission levels. Policy makers and governments worldwide are also increasingly demanding transparency and clarity from companies regarding their carbon emissions \cite{alsalemi2020achieving}. However, it is crucial to recognize that even the Information and Communication Technologies (ICT) deployed to enhance energy efficiency in buildings, such as cloud-based data centers and networks, could contribute to more than 14\% of global emissions by 2040 \cite{koot2021usage,himeur2023ai}.
Moreover, it is projected that data centers will account for over 33\% of the ICT industry’s global electricity consumption by 2025 \cite{koronen2020data, Edge-stats}. This creates an urgency to explore energy-efficient computing alternatives that can provide the computational resources necessary for ICT-based energy-saving solutions. Edge computing stands out as a significant alternative \cite{elnour2022neural}. It could serve as an optimal choice for energy-saving solutions in building management, significantly influencing the balance of energy consumption in a positive way \cite{kaur2018edge,sardianos2020rehab}.

One recommended strategy to reduce energy consumption in buildings involves the implementation of smart energy systems \cite{sayed2023time}. These systems largely hinge on the use of Internet of Things (IoT) and Artificial Intelligence (AI), converging into what's referred to as the Internet of Energy (IoE) technology \cite{joseph2020smart,pan2015internet}. IoE seeks to optimize the efficiency of energy generation, transmission, and usage. \textcolor{black}{IoE represents a transformative concept in the realm of energy management and distribution. Much like IoT, IoE connects devices, sensors, and infrastructure within the energy sector to create a highly interconnected and data-driven network. This mutual dependency enables real-time monitoring, control, and optimization of energy generation, distribution, and consumption. By harnessing advanced data analytics and automation, IoE promises to enhance energy efficiency, reliability, and sustainability while facilitating the integration of renewable energy sources and empowering consumers to make more informed energy choices. Ultimately, IoE has the potential to revolutionize how we generate, distribute, and interact with energy, steering in a more intelligent and responsive energy ecosystem. IoE leverages IoT devices to create intelligent sensor networks, opening the door to a range of energy-related applications. These include energy anomaly detection \cite{himeur2020novel,himeur2020smart}, energy usage optimization \cite{nizami2019multiagent}, energy prediction \cite{luo2019short}, non-technical loss (NTL) detection \cite{sharma2021unsupervised,bazau2019detection}, and non-intrusive load monitoring (NILM) \cite{himeur2021intelligent}, among others.}

%C. why Edge artificial intelligence is important\\

Artificial Intelligence (AI) is instrumental in devising efficient energy-saving solutions. It plays a crucial role in the monitoring, collection, control, evaluation, and management of power usage in households, public buildings, and industrial factories \cite{yu2020deep,bousbiat2023neural}. Furthermore, AI facilitates control over power usage, particularly during peak hours, and assists in the detection of consumption abnormalities, promptly notifying end-users \cite{himeur2022two}. It is also adept at identifying and predicting malfunctioning appliances and equipment \cite{han2020efficient}. With its capability to compress and analyze voluminous data, AI aids in the monitoring and interpretation of data generated by buildings and industrial environments, optimizing energy usage \cite{atalla2022recommendation}. It relies on big data analysis for decision-making, actively optimizing energy consumption and preemptively addressing issues based on predictive analysis \cite{wei2020deep}. Thus, AI-based solutions can process multi-modal data pertaining to energy usage, ambient conditions, and end-users' preferences to make informed predictions \cite{himeur2020data}. While this broadens the applicability of AI in IoE, it also necessitates high computational resources and trained problem solvers to tailor every AI-based solution, thereby addressing each unique problem \cite{sayed2022artificial}.

Cloud computing has been a pivotal component in the evolution of Artificial Intelligence (AI). It has unlocked a plethora of opportunities for deploying and advancing sustainable solutions without the need for investment in new hardware or software equipment \cite{alsalemi2020cloud,zhou2019edge}. Besides its data storage capabilities, cloud computing provides AI-based energy-efficiency solutions with robust analytics, comprehensive reports, personalized notifications and workflows \cite{bousbiat2023crossing,yu2022edge}. Furthermore, it can be integrated with other applications to collate data from multiple systems.
However, the escalating use of cloudlet platforms and networks contributes to significant bandwidth consumption, leading to increased energy use \cite{zhou2023profit,sayed2021endorsing}. This issue extends beyond energy services to various applications heavily reliant on networks and data centres \cite{alrazgan2022internet}. Moreover, other drawbacks compromise the application of cloud computing in IoE-based energy-saving solutions, such as dependency on network connection, loss of control, latency due to data transmission to the cloud, and concerns surrounding security and privacy \cite{kong2022edge,himeur2022techno}.

%---------------  Edge AI conceptual background and benefits
Edge AI, which refers to the local processing of AI algorithms on edge devices, can address many of the issues associated with cloud computing \cite{sayed2023edge}. Specifically, it negates the need to transmit data to the cloud for processing, enabling real-time decision-making and substantially reducing communication costs associated with cloudlet platforms \cite{taherizadeh2018monitoring}. In essence, edge AI brings processing and computational tasks closer to the point of interaction with the end-user, whether that be a smartphone, single board computer (SBC), domestic appliance, IoT device, or edge server \cite{alsalemi2021smart}.
Furthermore, edge AI provides a solution to the privacy concerns surrounding the transmission and storage of sensitive personal data in the cloud. It also mitigates bandwidth and latency issues that can limit data transmission capacity \cite{liu2019intelligent}. Consequently, edge technology has become essential not just for the energy sector but also for diverse fields such as autonomous vehicles, robotics, surveillance systems, healthcare monitoring, and diagnosis, among others \cite{kuo2019energy,kee2019non}.

\textcolor{black}{Nevertheless, large-scale data centers have the potential to be more energy-efficient due to economies of scale and the ability to invest in advanced technologies. However, achieving this efficiency requires careful planning, resource optimization, and a commitment to sustainable practices \cite{mahbod2022energy}. Smaller data centers can also be energy-efficient if they adopt best practices in design, operation, and energy management. Ultimately, the energy efficiency of a data center depends on a combination of factors, and size alone is not the sole determinant. Thus, optimizing those factors to achieve energy efficiency in large-scale data centers is not straightforward \cite{mahbod2022energy}.} 

\textcolor{black}{
Consequently, edge computing provides notable energy efficiency benefits compared to large data centers by reducing data transport, lowering latency, and enabling dynamic resource allocation. With processing occurring closer to data sources, there's less need for energy-intensive data transmission, leading to reduced energy consumption \cite{perez2021energy}. Edge devices can be designed for optimal energy use, powering down during idle periods. Furthermore, the decentralized nature of edge computing allows for localized renewable energy integration, improving overall sustainability. Load balancing and redundancy strategies in edge computing can enhance energy efficiency while ensuring resilience. Although the suitability of edge computing depends on specific use cases, its energy-saving potential makes it an attractive option for various applications \cite{patsias2023task}.} 

Looking ahead, edge AI eliminates the need for dedicated maintenance by data scientists or AI developers. Given that collected data are processed directly and delivered for monitoring, it becomes an autonomous technology \cite{}. Furthermore, there are no practical application limits for edge AI technology, giving it enormous potential for future development across a wide range of sectors \cite{himeur2022next}.

%Because of the Corona virus pandemic, the ingenuity of the companies has led to deploy solutions based on AI to provide accurate information in real-time. In healthcare, for example, AI is helping with patient monitoring, testing and treatment.

%---------------  Contribution of the paper
Given the significant benefits of edge AI, its market is projected to experience exponential growth in the near future, with the energy sector being a key driver of this expansion. Accordingly, this paper offers, to the best of our knowledge, the first comprehensive review of edge AI techniques for IoE applications. The key contributions of this paper are as follows:

\begin{itemize}[leftmargin=*]
\item The first review article that discusses the use of edge AI for IoE is introduced in this study.
\item A meticulously constructed classification of existing frameworks is initially carried out, examining various aspects integral to their development, including latency, real-time analysis, scalability, preservation of security and privacy, automated decision-making, and cost reduction;
\item We present an exploration of edge AI's primary applications within the IoE, underscoring its potential and the benefits it can bring to a range of use case scenarios, such as edge-based big energy data analysis, edge-based energy anomaly detection, edge-based NILM, edge non-technical loss detection, and edge-based energy recommender systems (RSs);
\item We conduct an extensive examination of the techniques utilized for swift inference, with in-depth descriptions of various strategies such as on-device computation, edge server computation, computing across edge devices, and private inference;
\item We carry out a critical analysis and discussion, outlining both resolved and unresolved challenges, as well as key insights drawn from the overview conducted; and
\item We offer insights into future perspectives and emerging areas of interest in both near and long-term research and development, including edge-to-edge cooperative AI, 5G/6G for edge AI, federated edge AI, reinforcement learning, and blockchain edge AI.
\end{itemize}

%---------------  Organisation of the paper

The structure of this paper is organized as follows: Section \ref{sec2} outlines the research methodology employed for this review and introduces the primary research questions addressed in this study. In Section \ref{sec3}, we discuss the technological requirements for edge AI in the context of IoE. Section \ref{sec4} offers an overview of edge AI techniques, structured within a comprehensive taxonomy. Section \ref{sec5} delves into techniques optimized for rapid inference, while Section \ref{sec6} examines contemporary methods for training AI on the edge. A thorough analysis and discussion take place in Section \ref{sec7}. Following this, Section \ref{sec8} presents emerging perspectives that are garnering significant research and development attention. The paper concludes with Section \ref{sec9}, where we summarize the key findings and concluding remarks.

\color{black}
\section{Review Methodology} \label{sec2}
\subsection{Objectives and Research Questions}
The aim of this research is to provide a comprehensive understanding of the role and benefits of edge AI techniques in the context of the IoE. By examining methods and applications of edge AI, the research seeks to highlight its advantages, such as real-time analytics, improved security, and cost-effectiveness. The study also critically evaluates the challenges currently facing edge AI, including security issues and computational constraints. %Additionally, the research anticipates future developments in this field, exploring emerging trends like the integration of 5G networks and blockchain. Through this review, the research intends to equip readers with a detailed insight into the significance, current status, and potential future of edge AI in IoE. 
Table \ref{RQs} presents the main identified research questions.
The selected research questions for this comprehensive review on the implications of Edge AI in the IoE serve to establish a strong foundation and provide a structured approach to examining this dynamic field. Beginning with the fundamental question of what Edge AI is and its significance in IoE, the review aims to clarify the core concept and highlight its advantages. Subsequently, it explores the practical applications and use cases of Edge AI within IoE, delving into the methodologies and strategies employed for rapid data analysis and decision-making. Additionally, the review seeks to identify and assess current challenges and areas necessitating further investigation in the realm of Edge AI for IoE. Lastly, it projects the future trajectory of this technology, anticipating technological advancements and trends that will shape the integration of Edge AI and IoE. Together, these research questions guide the review in offering a comprehensive understanding of Edge AI's role in IoE, from its foundational aspects to its future prospects.

%to achieve the objectives outlined above.
\begin{table*}[t!]
\caption{Research questions covered in this review.}
\label{RQs}
\small
\color{black}
\begin{tabular}{
m{8mm}
m{70mm}
m{70mm}
}
\hline

RQ\# & Question	& Objective   \\

\hline

RQ1 & What is Edge AI and its Significance in IoE and what are the primary benefits of implementing edge AI in IoE systems?	& 
Elucidate the concept of Edge AI and its relevance in IoE, as well as to identify the main advantages of utilizing edge AI within IoE systems.   \\ \hline

RQ2 & How is edge AI applied in IoE, and what technologies are used and what are the associated requirements ?	&  Explore the practical applications of edge AI in IoE and to highlight its main technologies and requirements.   \\ \hline

RQ3 & What techniques are employed to achieve fast inferences in edge AI-based IoE?	& Identify and understand the various methodologies and strategies used in edge AI to facilitate rapid data analysis and decision-making.   \\ \hline

%RQ4 & How is privacy maintained during inference in edge AI systems?	& Explore the mechanisms, strategies, and techniques implemented in edge AI systems to ensure data privacy during the inference process.   \\ \hline

%RQ5 & How is AI training conducted at the edge, and what are the parameters and protocols associated with it?	& Investigate the methodologies and processes used for AI training at the edge, and identify the specific parameters and communication protocols integral to such training.   \\ \hline

RQ4 & What are the current challenges, issues, and areas requiring critical analysis in the domain of edge AI for IoE?	& Identify, explore, and assess the existing challenges and concerns in the field of edge AI for IoE, and highlight areas that demand in-depth scrutiny and consideration for future advancements and solutions.  \\ \hline

RQ5 & How is the future of edge AI for IoE envisioned, and what new advancements and technologies are expected to shape it?	& Project and describe the potential future trajectory of edge AI for IoE, identify anticipated technological breakthroughs and trends, and understand how these advancements will influence and transform the landscape of IoE and edge AI integration.   \\ \hline

\hline

\end{tabular}
\end{table*}

\subsection{Search Strategy}
\begin{figure}[ht]
\centering
\includegraphics[width=\linewidth]{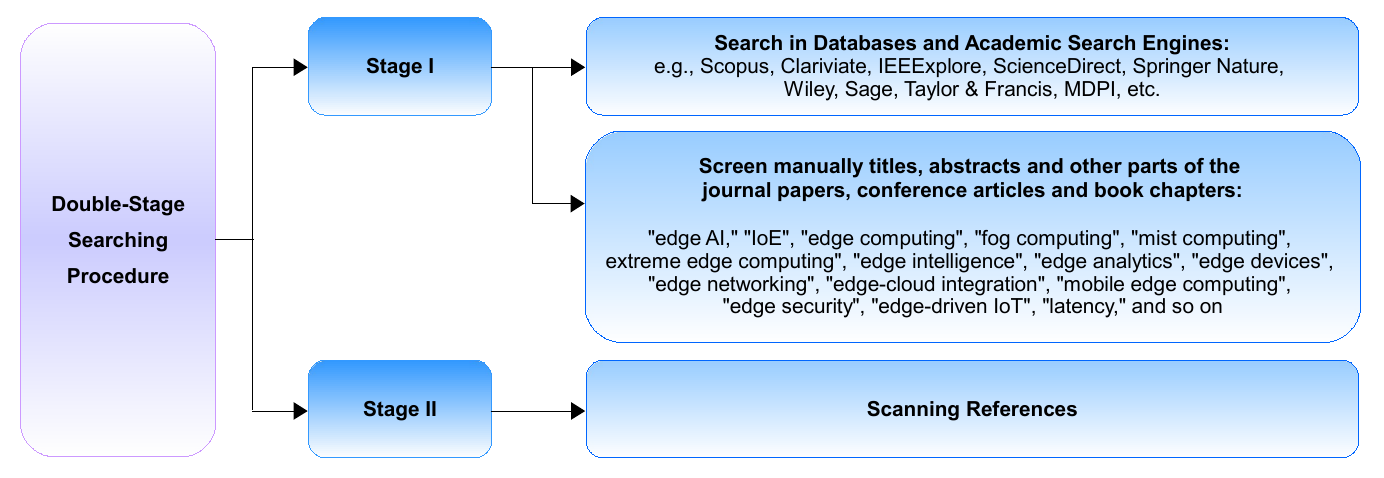}
\caption{\textcolor{black}{Implemented search procedure.}}
\label{search_procedure}
\end{figure}

To effectively gather literature for this review, a meticulous search strategy has been outlined. At the foundation of this strategy lies the identification of key terms, derived from the review's subtopics like "edge AI," "IoE", "edge computing", "fog computing", "mist computing", "extreme edge computing", "edge intelligence", "edge analytics", "edge devices", "edge networking", "edge-cloud integration", "mobile edge computing", "edge security", "edge-driven IoT", "latency," and so on. Leveraging Boolean operators, these keywords are systematically combined, producing search strings such as "Edge AI AND IoE", "Edge AI AND Latency", "fog computing AND IoE", "mist computing AND IoE", etc.
Subsequent to this, a selection of reputable databases becomes crucial. Resources like IEEE Xplore, ACM Digital Library, Springer Nature, Taylor \& Francis, Google Scholar, and ScienceDirect, among others, are scoured for relevant studies. In narrowing down the vast literature landscape, filters are applied. \textcolor{black}{Recent works, predominantly from 2017 onwards, are prioritized.} Furthermore, considerations are made to include only English language papers and those where full texts are accessible. Fig. \ref{search_procedure} provides an overview of the databases, libraries, and search engines that were included in our search strategy.

\subsection{Inclusion and Exclusion Selection Criteria}
During our comprehensive systematic review, we initially gathered a total of 512 research papers. To ensure that these papers were aligned with the focus of our study, we applied some fundamental screening criteria, including an assessment of titles, abstracts, and their relevance to our research question. Following this initial screening, we established more detailed inclusion and exclusion criteria to streamline our paper selection process.
Our inclusion criteria were designed to encompass papers that proposed solutions and addressed the specific keywords relevant to our research. Conversely, our exclusion criteria were applied to eliminate publications that did not directly address the topic of the IoE.
To maintain consistency and rigor in our paper selection, one of our team members took the lead in the selection strategy and conducted the initial screening. In cases where disagreements arose regarding the suitability of particular works, we resolved these through collaborative discussions involving all authors.
After removing any duplicate papers from our initial set, we identified a total of 337 unique articles. We then conducted a meticulous assessment of these remaining articles by carefully reviewing their titles, abstracts, and concluding sections. Based on this thorough assessment, we narrowed down our selection to 289 articles that demonstrated relevance primarily based on their titles and abstracts.

In the subsequent stage of our review process, we applied our specific exclusion criteria to the remaining articles, resulting in the exclusion of studies that did not meet our predetermined criteria.

\color{black}
Typically, during the selection process, we applied the following exclusion criteria to refine our literature set:
\begin{itemize}
\item Duplicate records
\item Articles from non-peer-reviewed sources.
\item Articles published before 2017.
\item Papers related to the implementation of applications utilizing previous RSs.
\item Papers related to research sectors other than IoE.
\item Papers written in languages other than English.
\end{itemize}

%\color{black}
%Furthermore, the following inclusion criteria were used to select relevant literature:
%\begin{itemize}
%\item Peer-reviewed journal articles, conference papers, and scholarly book chapters. 
%\item Articles that use a variety of research methodologies, including experimental studies, case studies, simulations and modeling.
%\item Articles published between 2017 and 2023.
%\end{itemize} 

\begin{figure}[!t]
\centering
\includegraphics[width=1\columnwidth]{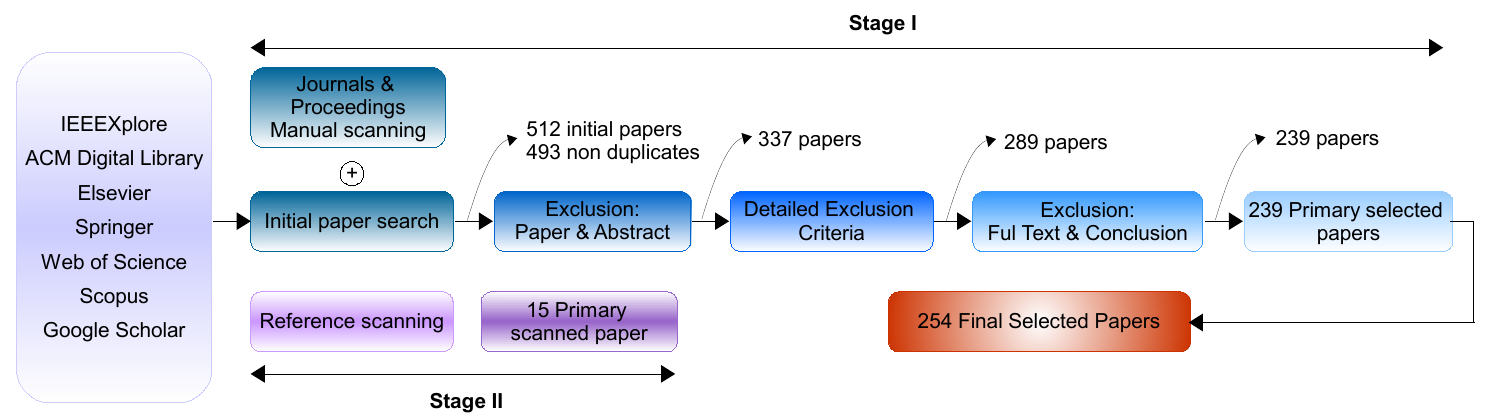}
\caption{\textcolor{black}{Study selection criteria.}}
\label{study_selection} 
\end{figure}

\textcolor{black}{
By applying these exclusion and inclusion criteria, we ensured that the selected articles provided insights, solutions, or advancements specifically related to the edge AI-based IoE applications. This process resulted in a final selection of 239 articles that met our inclusion criteria. In order to ensure a comprehensive review, we conducted a reference scan of the selected articles, which led us to identify an additional 15 relevant papers. Consequently, a total of 254 articles were included in our systematic review. Fig. \ref{study_selection} provides an overview of our research selection criteria and the distribution of publications obtained from each database.}

\color{black}

\section{Edge AI Technological Requirements} \label{sec3}
\textcolor{black}{Section 3 dives into the core technological requirements for Edge AI and mainly focuses on answering RQ2.}
Edge AI brings the prowess of artificial intelligence directly to the devices where data is generated, rather than relying solely on centralized cloud servers. This localized processing is made possible by an array of advanced technologies. Hardware components such as microcontrollers, ASICs, FPGAs, GPUs, and dedicated Neural Processing Units provide the necessary computational capabilities \cite{dou2023towards}. Software frameworks like TensorFlow Lite and PyTorch streamline the development and deployment of AI models tailored for edge devices, while operating systems, such as embedded Linux distributions ensure smooth operations. Networking technologies, encompassing both long-range options like LPWAN and short-range solutions like Bluetooth, facilitate seamless communication \cite{sjovall2023fpga}. Meanwhile, efficient power management mechanisms are paramount, given that many edge devices operate on limited energy sources \cite{ku2020state}. Moreover, in an age where data privacy and security are paramount, technologies like Secure Element Chips and Trusted Execution Environments are indispensable, ensuring that data remains protected at the edge \cite{tito2021edge}. The significance of Edge AI technologies cannot be overstated; they not only enable real-time data processing, enhancing responsiveness and efficiency, but also reduce data transmission costs, improve privacy, and extend battery life, thereby revolutionizing the landscape of AI deployments in myriad applications \cite{xu2021multi}. 
Below are some of the key technologies required:

\subsection{Hardware Components}
Implementing Edge AI is a multidisciplinary endeavor that requires a combination of expertise in AI, embedded systems, networking, and security. But as the demand for real-time, local data processing grows, so does the importance of Edge AI in various applications like smart cities, agriculture, healthcare, and manufacturing. Below are some of the key technologies required for implementing Edge AI:
\begin{itemize}
\item Microcontrollers (MCUs): For simpler AI tasks, modern microcontrollers can suffice. 
\begin{itemize}
\item ARM Cortex Series:
\begin{itemize}
\item Cortex-M: A series primarily used for real-time operating systems. Many AI platforms for edge devices, like TensorFlow Lite Micro, support Cortex-M MCUs. Examples include STM32 (from STMicroelectronics), nRF52 (from Nordic Semiconductor), and LPC (from NXP) \cite{lucan2022brief}.
\item Cortex-A: More powerful than Cortex-M, these are typically used in applications requiring higher computational capabilities. Examples include the Raspberry Pi, which uses Cortex-A-based Broadcom SoCs \cite{ray2022review}.
\item Cortex-R: Designed for real-time applications, though less common for AI compared to the other two.
\end{itemize}

\item Espressif Systems:
\begin{itemize}
\item ESP32: A popular low-cost MCU with Wi-Fi and Bluetooth capabilities. Its dual-core and ample memory make it suitable for some lightweight AI tasks \cite{arora2023modern}.
\end{itemize}

\item Microchip Technology:
\begin{itemize}
\item PIC32: Uses the MIPS architecture and can be used in embedded AI applications \cite{guimaraes2021optimization}.
\item AVR and SAM: These are other families from Microchip (originally from Atmel) that can handle lightweight AI tasks \cite{posch2019hands}.
\end{itemize}

\item RISC-V Architecture: While ARM is proprietary, RISC-V is an open-source architecture. Companies like SiFive offer RISC-V based MCUs, which are gaining traction in the embedded world, including AI at the edge \cite{talk2019game}.

\item Texas Instruments (TI):
\begin{itemize}
\item MSP430: Low-power MCU family suitable for battery-operated devices \cite{alsahli2022lightweight}.
\item Tiva C series: ARM Cortex-M based MCUs with a good balance of power efficiency and performance \cite{rupapara2023dynamic}.
\end{itemize}

\item NXP Semiconductors:
\begin{itemize}
\item i.MX RT series: Often referred to as crossover processors, they merge features of MCUs with those of more powerful application processors, making them suitable for more computation-intensive edge AI tasks \cite{oliveira2023investigating}.
\end{itemize}

\item Ambiq Micro:
\begin{itemize}
\item Apollo Series: Known for ultra-low power consumption, these ARM Cortex-M4F based MCUs are used in battery-critical edge AI applications \cite{shen2022big}.
\end{itemize}

\item GreenWaves Technologies:
\begin{itemize}
\item GAP8: A RISC-V based MCU explicitly designed for AI and high-performance computing on edge devices with tight power budgets \cite{lin2023imcu}.
\end{itemize}

\end{itemize}
\item Application-specific integrated circuits (ASICs): ASICs are integrated circuits tailored for a specific application, as opposed to general-purpose chips. Their design revolves around a specific task, and this focused nature allows for optimizations that can lead to superior performance, greater efficiency, and potentially reduced power consumption in comparison to more generic solutions \cite{petrou2022first}. In the context of AI, ASICs can be meticulously designed to handle specific neural network architectures or AI operations, resulting in rapid processing speeds and enhanced efficiency. For instance, Google's Tensor Processing Unit (TPU) is an ASIC crafted to accelerate TensorFlow-based machine learning workloads \cite{kaurbrar2022methodology}. By using ASICs, companies can achieve the desired balance between performance and power, making them invaluable for resource-constrained edge devices and high-performance computing environments. However, the trade-off is flexibility; while general-purpose processors can be repurposed or updated with software changes, ASICs are fixed in their functionality once manufactured \cite{jiang2023flexible}.
\item Field-programmable gate arrays (FPGAs): These are integrated circuits that can be customized post-manufactured to fit specific needs. They can accelerate AI workloads by enabling parallel processing \cite{azzouzi2023novel}. Their architecture consists of programmable logic blocks and reconfigurable interconnects, allowing for parallel processing—a boon for deep learning and neural network computations. FPGAs provide deterministic performance essential for real-time AI applications, ensuring consistent and predictable processing times. Additionally, their power efficiency makes them ideal for edge AI tasks with tight energy constraints. Supported by comprehensive toolchains from vendors and the advent of High-Level Synthesis, FPGA programming has become more accessible, solidifying their role as essential accelerators in the evolving AI landscape \cite{bobda2022future}.
\item Graphics Processing Units (GPUs): Originally designed for graphics, GPUs have proved to be highly effective for AI and deep learning tasks due to their parallel processing capabilities \cite{pandey2022transformational}. Boasting thousands of smaller cores capable of simultaneous processing and backed by high-bandwidth memory, GPUs are adept at tasks like matrix operations pivotal to neural networks. Their compatibility with leading deep learning frameworks, coupled with innovations such as Tensor Cores and dedicated AI hardware units, has further optimized AI computations \cite{hu2022survey}. Additionally, through frameworks like CUDA and OpenCL, GPUs have transitioned into general-purpose computing powerhouses. This computational agility, combined with the ability to link multiple GPUs, dramatically accelerates neural network training times, offering both economic and time efficiencies for researchers and businesses \cite{akter2023autism}.
\item Neural Processing Units (NPUs) or AI accelerators: Hardware specifically designed to accelerate neural network computations. These units elevate the concept of parallelism, inherently aligning it with neural network operations, and prioritize energy efficiency, making them indispensable for power-conscious edge devices and real-time AI applications like autonomous driving \cite{kim2023hardware}. Coupled with their adaptability and scalability features, NPUs effectively address memory bottlenecks by incorporating on-chip memory, ensuring swift data access. Their integration into comprehensive system-on-chip designs in modern devices like smartphones exemplifies their growing importance. Further cementing their pivotal role, manufacturers bolster NPUs with dedicated software ecosystems, enabling developers to fully exploit their capabilities, thus amplifying AI's reach and efficiency across diverse technological domains \cite{dhilleswararao2022efficient}.
\end{itemize}

\subsection{Software Components}

\begin{itemize}
\item AI Frameworks: AI frameworks, such as TensorFlow Lite, TensorFlow Micro, ONNX Runtime, and PyTorch, have revolutionized the deployment of AI models on edge devices \cite{sipola2022artificial}. TensorFlow Lite, a mobile-centric iteration from Google, and its even more compact counterpart, TensorFlow Micro, cater to devices ranging from smartphones to minimalistic IoT setups \cite{schizas2022tinyml}. Meanwhile, the ONNX Runtime offers a flexible, performance-tuned engine compatible with the universal ONNX format, suitable for diverse deployment platforms from cloud to edge. PyTorch, championed by Facebook for research, can seamlessly transition its models to edge deployment when converted to ONNX format. Collectively, these frameworks are reshaping the landscape, enabling efficient AI operations even on resource-constrained devices, and pushing the boundaries of real-time analytics and user interaction at the edge \cite{ray2022review}.
\item Operating Systems: Real-time operating systems (RTOS) like FreeRTOS, Zephyr, or embedded Linux distributions can handle the demands of Edge AI processing. Operating systems tailored for real-time and embedded applications, such as FreeRTOS, Zephyr, and embedded Linux distributions, play a pivotal role in Edge AI's landscape. These systems, designed for precision and efficiency, ensure that data is processed without delays, making them essential for applications demanding immediate response times, like autonomous vehicles or industrial controls \cite{sha2022recent}. FreeRTOS, recognized for its lean footprint, is a favored choice for devices in industrial and smart environments, while Zephyr's modern design emphasizes security and scalability, making it ideal for a range of IoT devices. Embedded Linux, on the other hand, provides a versatile environment suitable for more complex edge devices, offering a blend of flexibility and a rich feature set. Together, these operating systems form the foundational layer for Edge AI, facilitating reliable, efficient, and timely processing \cite{mezger2022survey}.
\item Model Optimization and Pruning Tools: These tools are used to reduce the size of AI models, so they can run efficiently on edge devices with limited resources.
\item Model Compilers: Convert models to an optimized format suitable for edge deployment, e.g., NVIDIA TensorRT, TVM, Glow \cite{shafi2022repercussions}. NVIDIA's TensorRT specializes in optimizing deep learning inferences for NVIDIA GPUs, exploiting reduced precision computations to enhance performance. TVM, an open-source offering, stands out for its vast device compatibility and its ability to auto-generate optimized code for diverse hardware \cite{lee2022quantune}. Meanwhile, Facebook's Glow enriches the landscape with its proficiency in graph optimization and hardware-agnostic design. Collectively, these compilers are paramount in reshaping machine learning models, ensuring swift, efficient, and adaptive deployments in Edge AI scenarios \cite{chen2023dycl}.
\end{itemize}

\subsection{Networking Technologies}
\begin{itemize}
\item Low-Power Wide-Area Networks (LPWAN) technologies, like LoRaWAN and NB-IoT, serve as critical pillars for efficient, long-range IoT communications. LoRaWAN, rooted in the proprietary LoRa modulation technique, offers expansive coverage and low power consumption, making it a preferred choice for applications such as smart agriculture and city infrastructure monitoring \cite{marini2022low}. In contrast, NB-IoT, a cellular-based approach, leverages a subset of the LTE standard to provide robust indoor coverage and reliability, supported by existing telecom infrastructure. Commonly used in smart metering and agriculture, both these LPWAN technologies ensure consistent, power-efficient, and long-distance data transmission, addressing the diverse needs of widely dispersed or remotely located IoT devices \cite{mousavi2022role}.
\item Short-Range Wireless: Technologies like Bluetooth Low Energy (BLE) and Zigbee can connect edge devices to local gateways or other devices.
\item 5G: The low latency and high speeds offered by 5G make it an enabling technology for edge AI, especially for applications that require real-time processing.
\end{itemize}

\subsection{Data Storage}
In the realm of Edge AI, where processing often happens in real-time and data might be generated in substantial quantities, having efficient storage solutions is paramount. Flash memory and small SSDs address this need, providing the perfect balance between speed, durability, and capacity. By temporarily holding data before it is processed, these storage mediums ensure that there's no lag or delay in operations, which is critical for applications like autonomous driving, real-time surveillance, and other responsive AI-driven systems \cite{monge2023ai}.

\subsection{Power Management Technologies}
In the expansive world of edge devices, especially battery-powered IoT sensors, power management and energy-harvesting technologies hold paramount importance. Efficient power management, facilitated by techniques such as Dynamic Voltage and Frequency Scaling and specialized Power Management ICs, ensures prolonged battery life and optimized energy consumption \cite{ben2022requirements}. These technologies deftly manage the energy needs of components, directing power selectively and transitioning devices into low-power states when they're not in active use. Furthermore, energy-harvesting mechanisms, like photovoltaic cells or thermoelectric generators, capture ambient energy from the environment, offering a supplementary power source that can, in some cases, sustain low-power devices indefinitely \cite{hasan2023study}. Together, these innovations empower edge devices to meet the stringent demands of real-time processing while conserving energy, especially vital for remotely located or inaccessible devices.

\subsection{Security Technologies}
\begin{itemize}
\item Secure Element Chips: they act as fortified vaults within devices, safeguarding sensitive data like encryption keys from potential breaches. Isolated from a device's primary hardware and software, these microprocessor chips are essential for tasks ranging from mobile payments to biometric data protection. Beyond software defenses, they offer robust physical security, resisting tampering and unauthorized access. With escalating cyber threats, their role in preserving digital trust and ensuring the sanctity of critical data, from credit card details to biometrics, is paramount \cite{zakaret2022blockchain}.
\item Trusted Execution Environments (TEE): Offers a secure area of the main processor where both code and data are protected \cite{lacoste2023trusted}. Additionally, they act as a shielded zone within a device's main processor, safeguarding sensitive operations and data from potential external threats. As the digital world continues to grow, and as threats become more sophisticated, the role of TEEs in ensuring the safety and privacy of user data and operations becomes increasingly vital \cite{ding2022roadmap}.

\item Hardware-backed Key Storage: bolsters the security of cryptographic keys by ensuring they are not only stored but also often processed within a secure hardware environment, isolated from conventional software vulnerabilities and attacks. As cyber threats become more advanced, such hardware-centric solutions are becoming indispensable in safeguarding sensitive digital assets \cite{plappert2022analysis}.
\item Remote Attestation: is a security protocol ensuring the authenticity and integrity of a device's software and hardware before initiating communication. In this process, the device, termed the "prover," measures its current state, which is then verified against a known baseline by another entity, the "verifier" \cite{menetrey2022watz}. Primarily used in IoT devices, cloud computing, and mobile devices, RA acts as a critical digital "trust handshake," ensuring devices operate in a secure, unaltered state. The practice plays a pivotal role in the modern digital landscape, addressing the increasing sophistication of cyber threats \cite{walther2022ratls}.
\end{itemize}

\subsection{Federated Learning (FL) Tools}
These are specialized tools and platforms designed for training AI models across decentralized devices, where the models learn locally, and only the model updates are aggregated centrally. FL is a groundbreaking approach to training AI models that prioritizes data privacy and efficiency, especially pertinent in today's data-driven world \cite{bousbiat2023crossing}. 
Unlike traditional centralized machine learning where all the data is sent to a central server for model training, FL pushes the model to the edge device (like smartphones or IoT devices) where the data resides \cite{varlamis2022using}.
In this approach, individual devices compute model updates locally on their own data. This means raw data never leaves the device, which is a significant step for preserving user privacy.

After local training on devices, only the model updates (e.g., weight changes) are sent back to a central server.
These updates from multiple devices are aggregated on this central server to produce a global model that benefits from the diverse data of all participating devices. This aggregation can be done using methods that ensure even the updates do not leak sensitive information.
Since raw data does not leave the device, the user's information is better protected against breaches or misuse.
Efficiency: Sending only model updates instead of large datasets reduces the need for significant bandwidth and central storage.
Real-world Relevance: Models can be tailored to real-world user data without violating the user's privacy, making them more effective in practical applications.

\subsection{Deployment and Monitoring Tools}
Platforms that aid in the deployment of AI models to edge devices, and subsequently, monitor their performance and health in real-time.
Accordingly, deployment tools streamline the process of transferring and installing AI models onto edge devices. This involves optimizing models for the specific constraints of edge devices, like limited computation and memory. Monitoring tools, on the other hand, ensure that once the AI models are deployed, they run efficiently, remain up-to-date, and any potential issues are detected and addressed in real-time.
%Implementing Edge AI is a multidisciplinary endeavor that requires a combination of expertise in AI, embedded systems, networking, and security. But as the demand for real-time, local data processing grows, so does the importance of Edge AI in various applications like smart cities, agriculture, healthcare, and manufacturing.

\subsubsection{Key features}
Key features include (i) model optimization: before deployment, the AI models often need to be trimmed, quantized, or pruned to fit the constraints of edge devices. Tools should offer capabilities for these optimizations without a significant loss in model accuracy; (ii) over-the-air (OTA) updates: as models get updated or improved over time, deployment platforms should facilitate seamless OTA updates to edge devices without causing downtime; (iii) real-time performance metrics: monitoring tools should provide insights into the model's inference speed, accuracy, and resource usage (CPU, memory, battery) in real-time; (iv) health checks: the tools should periodically check the health of the deployed models and ensure they are running without errors. If anomalies are detected, they should be flagged immediately; (v) logging and reporting: deployment platforms often include logging features, keeping a record of deployments, updates, and potential issues. These logs can be crucial for troubleshooting and performance tuning; (vi) scalability: tools should be able to handle deployments to a large number of devices simultaneously, especially in scenarios with vast IoT networks; and (vii) security features: ensuring secure model deployments, these tools should include encryption methods for data in transit and at rest, as well as secure authentication mechanisms.

\subsubsection{Popular Deployment and Monitoring Platforms}
Popular deployment and monitoring platforms inlcude (i) NVIDIA DeepStream: designed for deploying AI in video analytics on edge devices, offering tools for optimization and monitoring; (ii) AWS Greengrass: Amazon's platform for IoT deployment that provides ML inference capabilities at the edge, along with robust monitoring solutions; (iii) Azure IoT edge: Microsoft's platform facilitating deployment of cloud workloads to IoT edge devices, complemented by monitoring capabilities; and (iv) Google Cloud IoT edge: Provides capabilities to deploy and run ML models on edge devices while ensuring constant monitoring and updates \cite{singh2023edge,yousuf2022ibug}.

\section{Overview of edge AI techniques for IoE} \label{sec4}
\begin{figure}[!t]
\centering
\includegraphics[width=1\columnwidth]{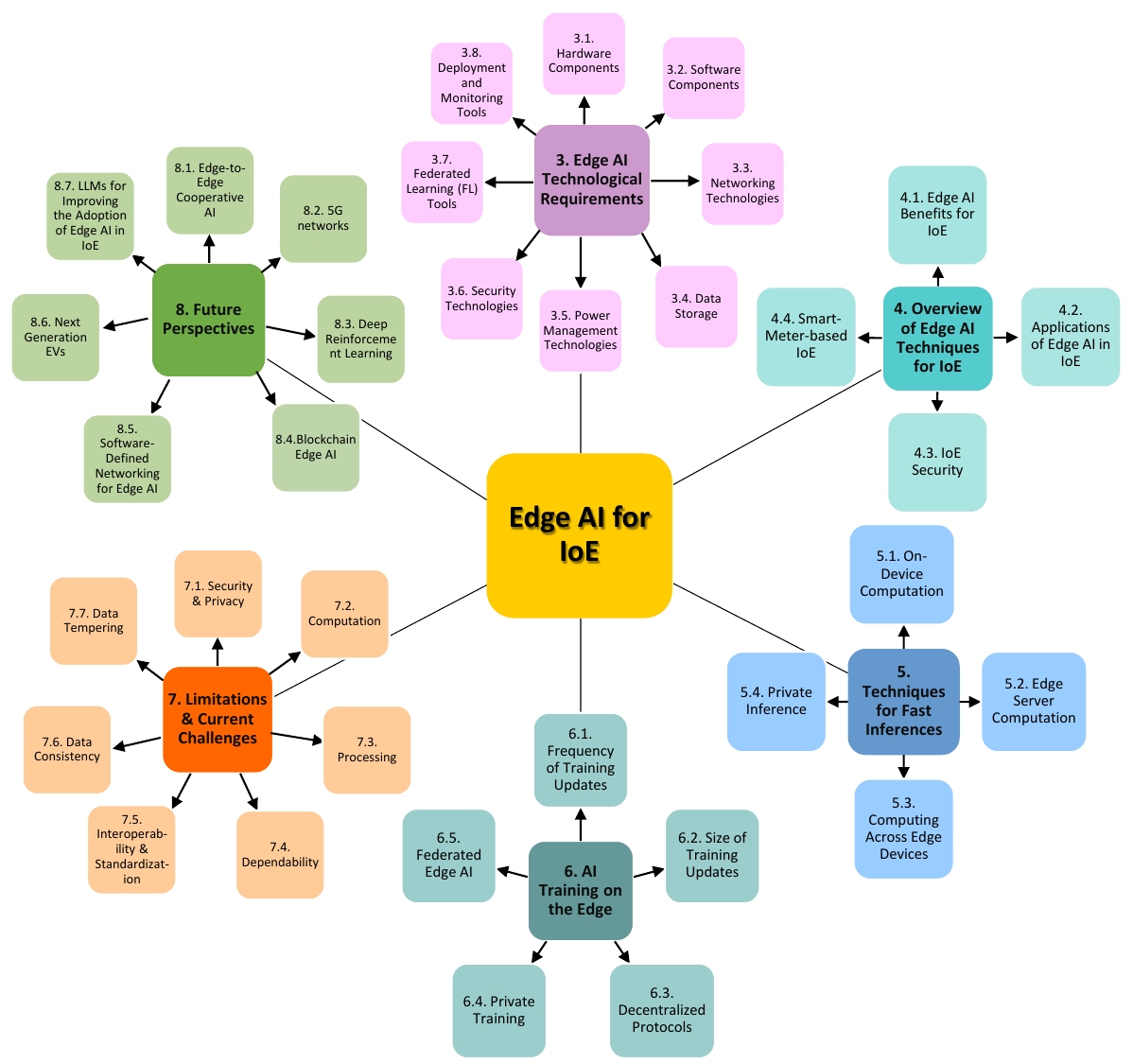}
\caption{Taxonomy of existing Edge AI techniques for IoE applications.}
\label{taxonomy} 
\end{figure}

\textcolor{black}{RQ3 is addressed in depth in this section, providing a comprehensive discussion of the techniques utilized for fast inference in edge AI-based IoE.}
To overview existing edge AI techniques for IoE, a well-defined taxonomy is adopted in this section, as it is portrayed in Fig. \ref{taxonomy}.

\subsection{Edge AI benefits for IoE} 

The conventional approach of energy integration and operating cost optimization employs cloud infrastructure technologies to centralize computing activities, raising the strain of computing \cite{fang2020edge}. Owing to the advancement of digital networking technologies, such as IoT, IoE and 5G networks, modern computer technologies are an important means of unloading technical functions on the edge on these networks. In addition, with the rise of the generated data, it is becoming a crucial point to efficiently boost the computational capacity of edge nodes in edge computing \cite{sardianos2020data}. Therefore, edge AI helps in speeding up decision-making, enabling data processing to be more secure, improving users' experiences with hyper-personalization, and lowering the cost through avoiding costly cloud services and boosting energy efficiency of devices \cite{himeur2021appliance,deng2020edge}. Moreover, edge computing has been recently recognized as a high-potential technology owing to its benefits in delivering low-latency computing services for energy mobile consumer devices and IoT applications \cite{zhang2019masm}.

%An example of this could be a hand-held tool used in a factory. The tool is embedded with a microprocessor that utilizes Edge AI software. The tool's battery lasts longer, when data doesn't have to be sent to the cloud. The tool collects, processes, and analyses data in real-time, and after the work day, the tool sends the data to the cloud for later analysis. A tool embedded with AI could for example turn itself off in the event of an emergency. The manufacturer receives valuable information about how their products are working and can utilize this information in further product development.

Fig. \ref{edge-IoE-exp} portrays an example of an edge-based IoE ecosystem proposed in \cite{alsalemi2021micro,himeur2021smart} to promote energy saving in buildings, which is part of the (EM)$^{3}$ project \footnote{\url{http://em3.qu.edu.qa/}}. The latter focuses on developing effective edge-based  energy efficiency solutions using AI, micro-moment analysis and behavioral change.

\begin{figure}[!t]
\centering
\includegraphics[width=1\columnwidth]{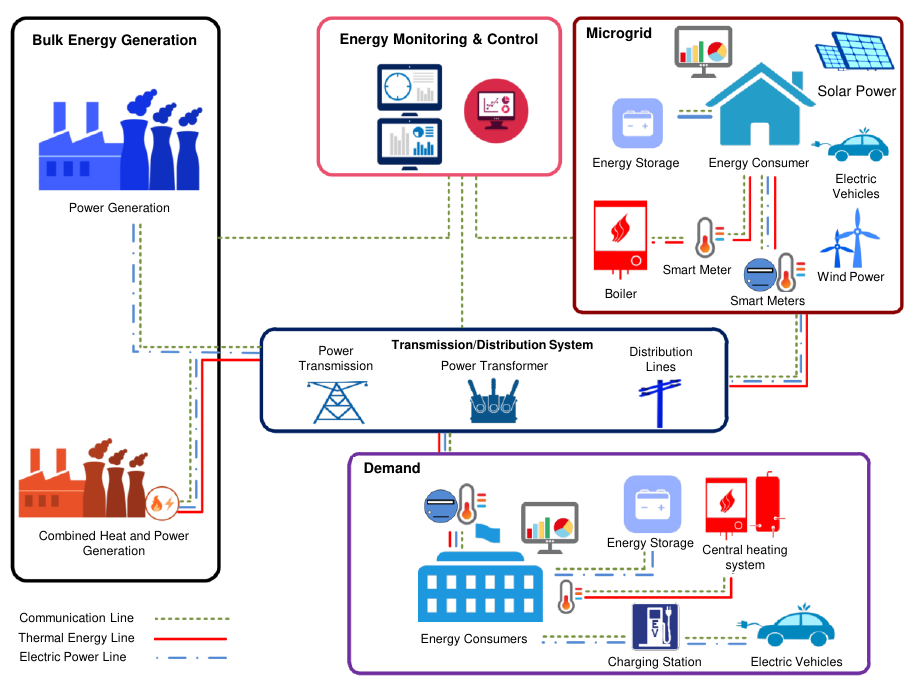}
\caption{Overall architecture of the IoE ecosystem.}
\label{edge-IoE-exp} 
\end{figure}

\begin{figure}[!t]
\centering
\includegraphics[width=1\columnwidth]{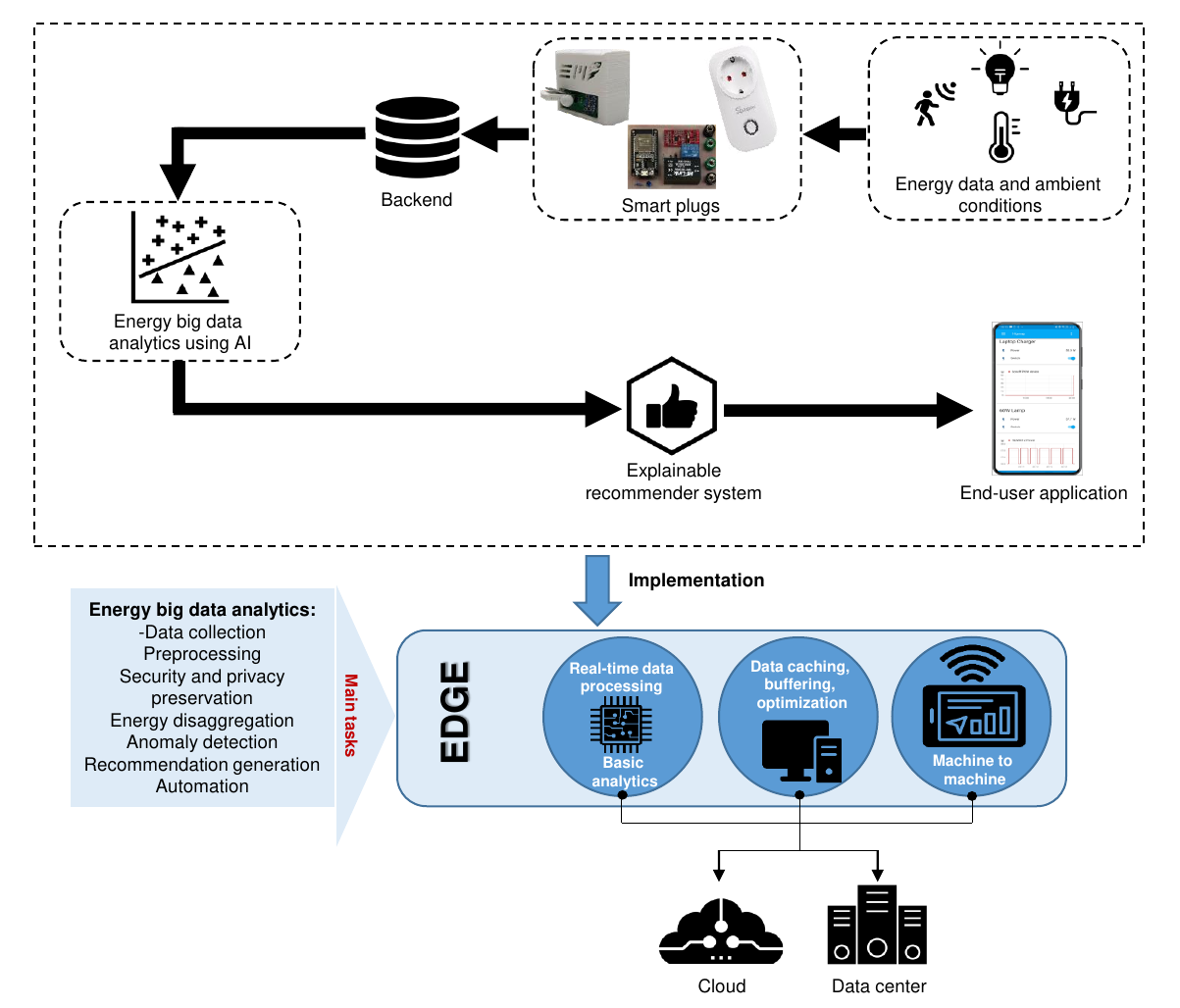}
\caption{Flowchart of the edge-based IoE ecosystem, called EM$^3$ proposed in \cite{himeur2022techno} for promoting energy saving in buildings. }
\label{edge-IoE-exp} 
\end{figure}

The Internet of Energy (IoE) involves distributed cognitive knowledge for data analysis and energy system decision-making, aiming for distributed, flexible, and privacy-friendly smart energy management in IoE \cite{zhong2019admm}. Firstly, clustered edge servers are deployed at the core network's edge, providing low-delay data storage and processing capabilities. Secondly, intelligent energy systems facilitate automated data collection, decision-making, and actuation.
In this context, we further explore the IoE enablers resulting from the adoption of edge AI. These enablers encompass reduced latency, real-time analytics, scalability, information security and privacy, automated decision-making, and cost-effectiveness

\vskip2mm
\subsubsection{Latency} 
To reduce the latency of cloud storage resources, in recent years, edge computing has evolved as an innovative strategy that can offer low-latency computing services for offloading. In edge computing, the Cloudlet server may be installed in the local access network to eliminate long-haul data transfer from consumers to the central cloud center, minimizing data transmission time dramatically \cite{premsankar2018edge}. In the other hand, by unloading computing intensive tasks to the Cloudlet, the operational energy usage in the device terminals may also be substantially decreased. In addition, edge computing can be used to include information caching and storage capabilities that help minimize network traffic and the user's perceived pause in data delivery \cite{zhang2018mobile}.

\vskip2mm
\subsubsection{Real-time analytics} 
Conducting rela-time or near real-time analytics may be achieved with edge computing, where the computation task takes place within a fraction of a second-which is important in time-critical conditions \cite{nastic2017serverless,trinks2018edge}. For instance, let us consider an equipment on a production line of the plant, if the robot on the production line is triggered at a wrong moment or too late, the product may be destroyed or the product may pass on to the assembly line unprocessed and unused. If the error goes overlooked, the defective product can end up on the market or trigger harm in later phases of manufacturing \cite{khan2019edge,yu2017survey}.

Devices are able to communicate over global networks, e.g. the Internet, where operators may examine real-time data obtained from complex energy supply, storage and utilization processes, however, edge data processing allows operators to react rapidly to complex developments, improve energy flow control and deliver useful services to end-users of energy, achieving real-time performance \cite{zhong2019admm,barthelemy2019edge}.

\subsubsection{Scalability} 
Research company IDC forecasts that there would be 41.6 billion linked IoT devices producing 79.4 zettabytes of data in 2025 \cite{IoT-IDC}. As volumes rise, new creative approaches for efficient analysis and big data processing are required. 
When much data processing is performed locally, consolidated operation or data sharing at the edge would not result in bottleneck circumstances. Typically, edge AI applications require vast volumes of data. For example, if the data's nature is video or image data from hundreds or thousands of separate sources at the same time, moving data to a cloud server is not a feasible option.
Despite the promising potential, according to \cite{rausch2019towards}, handling and orchestrating cloud-based edge services is more complicated since nodes usually lie behind private networks or firewalls. Further contributions are necessary to prove the scalability of edge AI solutions.

\subsubsection{Information security and privacy} 
Less data in the cloud means fewer chances for online threats. Edge also runs in a closed network, making it difficult to hack details. It is much tougher to pull down a network made up of several computers \cite{xiong2019ai}. Generally, one might assume that something that has a protection aspect needs to be handled on the edge. As an illustration, we may think of the intelligent safety control systems in the warehouse. When computers are not functioning as they should be, even when individuals are going through a restricted place, the warning should be turned off until the accident has occurred \cite{zhang2019differential}. 
As already stated, where data processing takes place locally, there is no need to transfer data to a cloud environment. This makes it pretty tough to access data without authorization. Even, critical data that is stored in real-time, such as video data, can only live for a blink of an eye until it disappears. In such cases, it is better to guarantee data protection and confidentiality since the attacker may have clear access to the actual computer where the data is being stored.

Edge computing has the potential for consumer security and privacy, which is said to boost the existing state of the art in these fields \cite{sachdev2020towards}. For example, data processing can be performed locally, and protection and privacy can be improved. However, in such a setting, edge AI can be vulnerable to its own protection and privacy concerns, especially in the context of digital marketing where personal data is involved. The continuing task is to ensure confidentiality in this sense and to satisfy the numerous legal standards of privacy while they continue to change and many of which are not completely apparent from a technological point of view.

\subsubsection{Automated decision-making} 
The paradigm of edge IoE can build intelligent stand-alone goods by forging efficient sensing, self-learning, wisdom as a service (WaaS), information as a service (InaaS), precise decision-making and actuation using successful location-independent tracking, control and inventory management strategies, as well as retaining a competitive edge by increased product efficiency by immediate and enhanced product performance \cite{kuru2019transformation}.

%The confluence of mechanical systems and microelectronics opens up fresh possibilities for mechanical modeling and automated operations together with knowledge-based systems and learning capabilities. Microcontrollers are gradually integrated in electromechanical systems \cite{kuru2019transformation}, providing even more precision and control capabilities in the architecture of the machine.

%For example, there is a multitude of sensors in a self-driving car that continually tracks, e.g. the direction of the vehicle and the speed of rotation of the tyre. The driving machine automatically makes the required decisions about steering, braking and the usage of the throttle based on the data obtained from the sensors.

\subsubsection{Reduced costs}
Edge computing offers notable cost savings by optimizing the use of cloud computing architectures, ensuring faster analytics, and reducing decision-making lag. It conserves bandwidth and reduces the need for data transmission, leading to decreased computing costs \cite{yousefpour2019all}. While cloud data processing can be expensive, especially when large bandwidth is required, edge AI, despite its local processing resource costs, is typically the most economical choice. This allows IoE providers to monitor and manage their smart devices in real-time, enhancing their reliability and cost-effectiveness \cite{kuru2019transformation}. In \cite{zhou2020iot}, a cloud-edge resource provisioning system is introduced, focusing on delay-aware Lyapunov optimization. Despite not having prior knowledge of cloud-edge device statistics, the system can make informed online decisions about resource allocations for diverse IoE applications, proving its cost-effectiveness through empirical analysis.

\subsection{Applications of edge AI in IoE (A)}
Distributed energy resources generate over 1 TB of data daily, which centralized cloud platforms manage. As assets produce data exceeding what infrastructure can efficiently transfer, the centralized cloud system may face latency and cybersecurity issues \cite{EdgeComputeCleantech}. Edge computing addresses these challenges by processing data on distributed nodes, leading to faster processing and improved security. While batch uploads are sent at regular intervals for pattern analysis, edge data transmission rates can surpass cloud computing by up to 1000 times, impacting AI-based learning and advanced analytics \cite{khan2020edge}. Critical infrastructures often limit access due to cybersecurity risks, but edge computing can bypass these by reducing reliance on internet connections. By 2021, enterprise-generated data stored at the edge is projected to rise to 75\% from under 10\% in 2018 \cite{ianculescu2019IoHT}. Edge computing, although not a new concept, offers innovative solutions, addressing challenges like reducing energy consumption of big data centers, ensuring data privacy, cutting costs, and enabling faster applications \cite{EdgeCapacityChipestimate}.

\subsubsection{Edge-based energy data analytics (A1)}
\begin{figure}[!t]
\centering
\includegraphics[width=1\columnwidth]{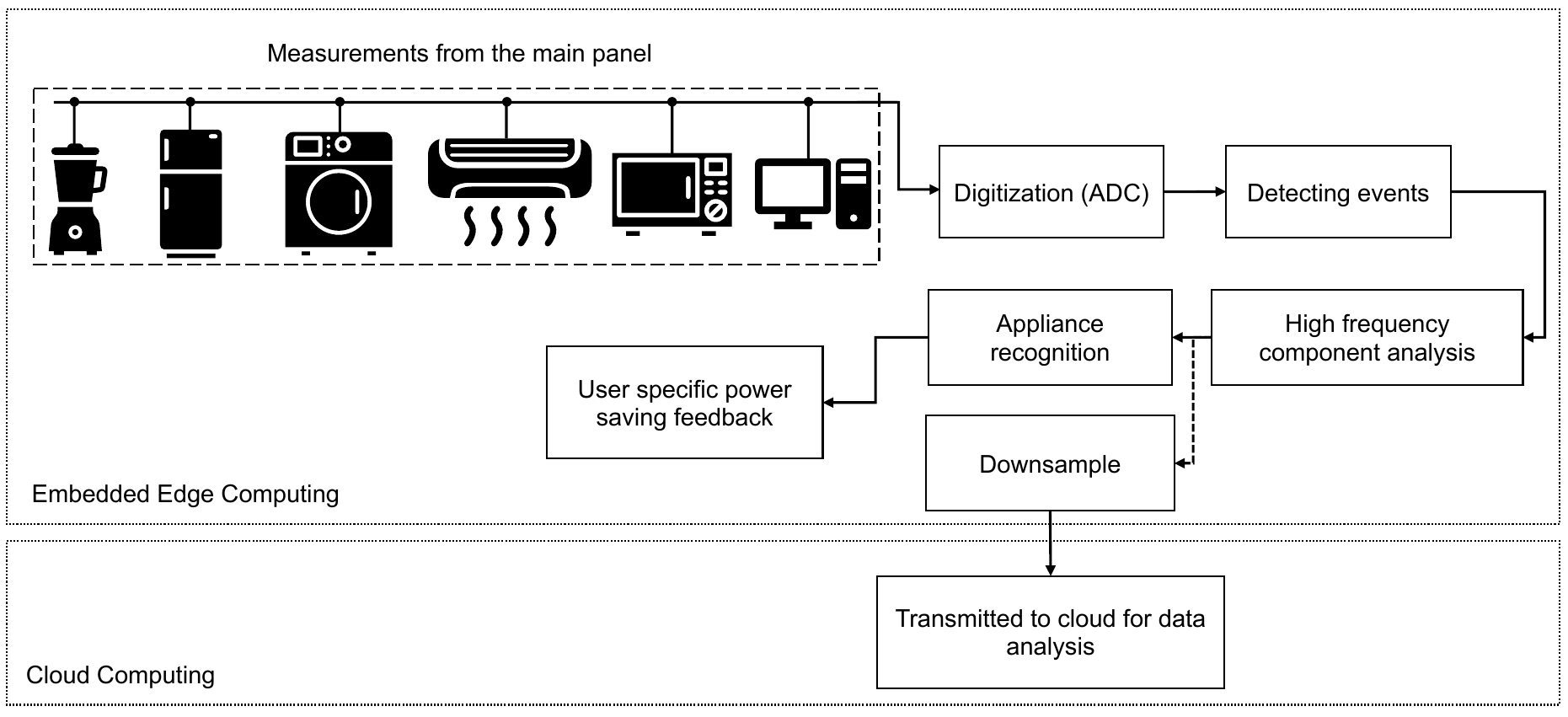}
\caption{Example of an edge smart meter data analysis system \cite{sirojan2019embedded}.}
\label{sm-data-analysis} 
\end{figure}

Cloud platforms, while beneficial for aggregating computing and storage needs, are not energy-efficient, consuming over 3\% of global electricity and emitting 2\% of CO2 emissions globally \cite{chang2018insight}. Their energy demand is set to grow significantly by 2030. These platforms often run continuously, even when not in use. Edge AI addresses this by efficiently managing energy demands, such as putting resources into sleep mode when not required \cite{zhou2018begin,garg2019edge}. Furthermore, edge AI's proximity to end-users allows for tailored configurations, making management of distributed edge devices easier and more energy-efficient than large cloud data centers \cite{sitton2019edge,singh2019fog}.
In smart grid advancements, traditional electric meters are being replaced by smart meters for enhanced accuracy and trend visibility. Although mainly used for remote monitoring, smart meter data holds potential for diverse applications like demand management and anomaly detection \cite{varlamis2020bds}. However, these aren't widely implemented due to constraints like high data sampling rates and bandwidth. Sirojan et al. highlights an edge computing framework's advantages in addressing these issues by bringing data analytics to smart meters, resulting in improved performance and efficiency for smart grids \cite{sirojan2019embedded}. An example of such an architecture is showcased in Fig. \ref{sm-data-analysis} \cite{samie2019edge}.

While various data analytical algorithms exist for processing large data sets, many fall short in execution. Research on streaming data analysis is ongoing to make methods more cost-effective. In \cite{ahuja2019data}, a data analytics platform tailored for smart energy meters is introduced, employing cutting-edge analytical tools combined with gamification to boost user engagement. This platform aims to benefit clients, utilities, and stakeholders by highlighting the advantages of analytics for smart energy meters. The proposed algorithm, enhanced by gamification, allows for real-time customer involvement, aiding researchers and authorities in future planning.

\vskip2mm
\subsubsection{Edge-based energy anomaly detection (A2)}
The proliferation of IoT has propelled the energy sector towards AI-driven anomaly detection at the edge, prevalent in diverse areas like smart cities and agriculture \cite{himeur2021artificial,himeur2020detection}. Edge solutions tackle issues like limited computational power and the inefficiencies of centralized big data due to communication costs, ensuring swift actions with minimal delay \cite{himeur2020emergence}. Anomaly detection, vital for sectors like energy and cybersecurity, has traditionally been managed centrally. Yet, the rise of edge computing now emphasizes on-device analytics \cite{himeur2021smartb}. For example, \cite{schneible2017anomaly} developed an edge-focused anomaly detection technique using autoencoders, minimizing data transmission by focusing on crucial data. Although IoE data analysis is critical for tasks like predictive maintenance, achieving accurate anomaly detection remains a challenge \cite{mohamudally2018building}. The evolution of edge-based Anomaly Detection Engines (ADE) underscores the complexities of IoT networks, highlighting the practicalities and difficulties of implementing ADE on the edge.

Distributed surveillance systems for anomaly detection have pivoted from centralized to edge computing due to real-time processing and bandwidth concerns \cite{sardianos2019reshaping}. Algorithms tailored for limited computing at sensors are emerging, with \cite{marchioni2020subspace} proposing lightweight techniques that match the efficiency of complex spectral analysis, apt for edge computing. Cloud computing's ability to handle cyber threats is contrasted with edge computing's advantages by \cite{xu2019data}, introducing a data-driven edge intelligence method for network anomaly detection. In the context of Vehicles Internet (VI), \cite{zhu2019mobile} introduces an edge-based LSTM framework for anomaly detection in the Internet of vehicles (IoVs) with a 90\% accuracy. Meanwhile, \cite{shah2016edgecentric} delves into anomaly detection in large edge-associated graphs, presenting EdgeCentric which detects unusual user behaviors in networks like Flipkart with 0.87\% accuracy. Similarly, the M2SP-EdgeIoE system proposed in \cite{alsalemi2022innovative}, rooted in Edge Computing, offers an innovative approach to enhancing home energy efficiency in IoE contexts. It comprises four core modules: data collection, appliance identification through energy patterns, AI-based anomaly detection in energy use, and a personalized recommendation engine. The system's effectiveness is underscored by its high accuracy rates of 98.49\% in recognizing appliances and 95\% in pinpointing anomalies. Overall, M2SP-EdgeIoE showcases the potential of Edge IoE in driving global energy-saving solutions.

\begin{figure}[!t]
\centering
\includegraphics[width=1\columnwidth]{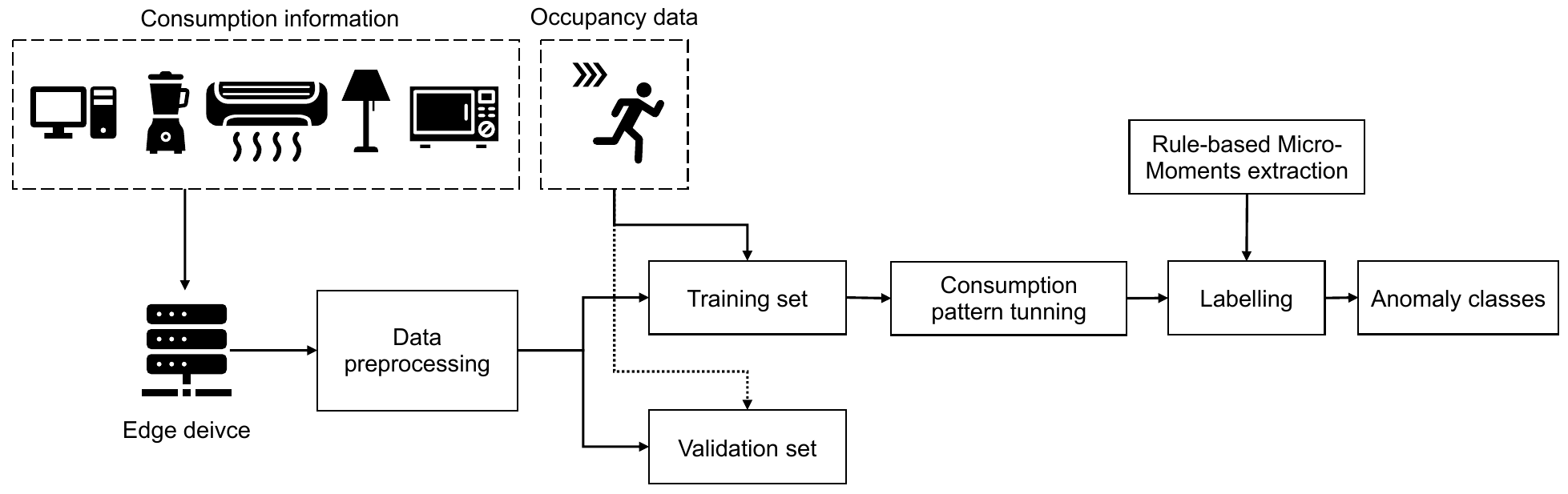}
\caption{Example of an edge anomaly detection approach utilizing the concept of micro-moments \cite{himeur2020novel}.}
\label{edge-anomaly-detection} 
\end{figure}

In \cite{hussain2019mobile}, deep learning is used to detect network anomalies like sleeping cells and high traffic using the mobile edge computing (MEC) model. With computations distributed across MEC servers near base stations, the system achieves a 98.8\% accuracy and a 0.44\% false positive rate (FPR), marking an advancement from previous models.
\cite{luo2018arrays} presents a fast, edge-based anomaly detection method using arrays of count estimators (ACE). This method, powered by a 3.50 GHz core Xeon platform, offers notable speed and robust privacy features. The ACE algorithm, rooted in locality sensitive hashing (LSH), surpasses 11 standard benchmarks, including the KDD-Cup99 dataset.
\cite{ngo2020adaptive} proposes an adaptive anomaly detection solution for hierarchical edge computing (HEC) systems, with three tailored DNN models for its various layers. Using contextual input data, the model selection process optimally balances detection accuracy and speed.
In \cite{lin2019edge}, an edge-based recurrent neural network (RNN) platform, ERADP, is introduced. To address the challenge of limited run-to-failure data, it adopts a reconstruction-based method to craft a secure system model. This model aims for a 100\% true alarm rate in anomaly detection and can speed up model training by up to 120 times.
\cite{ezeme2019deep} proposes a modular deep learning architecture combined with an offloading algorithm that taps into the increasing capabilities of edge devices. This creates a distributed anomaly detection system endowing every network node with real-time anomaly detection. When tested on kernel event streams, the results confirmed the algorithm's effectiveness, even under varying timing constraints.
Lastly, \cite{himeur2020novel} focuses on leveraging micro-moments extracted from energy consumption data, gathered from building sensors and utility sub-meters. These signals aim to detect unusual usage by end-users, taking into consideration their occupancy behaviors. The micro-moments are derived from a myriad of raw sensor signals, which then inform the subsequent fieldwork. The overall flowchart of the proposed method is shown in Fig. \ref{edge-anomaly-detection}.

% \cite{luo2020anomaly}

\vskip2mm
\subsubsection{Edge-based NILM (A3)}
Edge AI enhances smart grids for optimized energy management by connecting sensors and IoT devices to edge platforms in buildings for real-time energy monitoring \cite{himeur2020applicability}. Typically, \cite{tabanelli2020feature} focuses on real-time Non-Intrusive Load Monitoring (NILM) on resource-limited smart meters. A survey on feature frequencies enables NILM algorithm implementation on edge devices. Their MCU-based smart meter captures readings and performs NILM on the edge, achieving a 95.99\% accuracy using minimal memory and computational resources.
Domestic NILM typically involves cloud processing, but this approach raises concerns about cost and data privacy \cite{ahmed2020edge}. This research explores executing NILM algorithms on end-user devices, like mobile phones. A two-stage model was developed for efficient disaggregation: first, utilizing the MobileNet deep learning algorithm for a compact model, and then compressing it using TensorFlow Lite to minimize edge computing resources. The model was trained on a combination of HES, UKDALE, and REFIT datasets to address real-life variations.

Fog computing, an extension of cloud computing to the network's edge, offers reduced latency benefits, especially in smart grids for fine-grained energy data collection \cite{cao2019achieving}. This data helps in modeling power production but risks revealing user habits. NILM, which monitors individual appliance usage, can expose this privacy further. To protect user privacy, a privacy-preserving method is introduced for electrical load tracking. This method is based on the Factorial Hidden Markov Model (FHMM), and uniquely adds noise to the activity parameter, rather than the conventional approach of noise addition to energy data.
Earlier NILM research typically favored cloud-computing solutions, resulting in reduced sampling rates to conserve bandwidth, which compromised load detection accuracy \cite{hernandez2020design}. This study presents an SoC FPGA architecture, designed for local installation in homes or buildings. It manages high-frequency data sampling and real-time load detection using specific algorithms. Experimental results have initially validated this design.

The paper \cite{xiang2019iot} introduces Grid Sense, an innovative IoT-based structure for rapid load management, edge computing, and NILM. Grid Sense offers a cost-effective solution for large-scale individual load management using existing connectivity. As Grid Sense is now being used in several pilot projects by the State Grid Jiangsu Electric Power Company, this paper details its technology, system requirements, and performance results.
In another study \cite{liu2020secure}, an edge gateway is designed for secure load monitoring in smart energy systems. It presents a method to separate loads based on events from low-frequency signals, producing consumption profiles of various devices without requiring their primary data. After extracting key features from the collective signal, an appliance profile recognition technique is proposed. The efficiency of this method is proven using a public data-set.

\begin{figure}[!t]
\centering
\includegraphics[width=1\columnwidth]{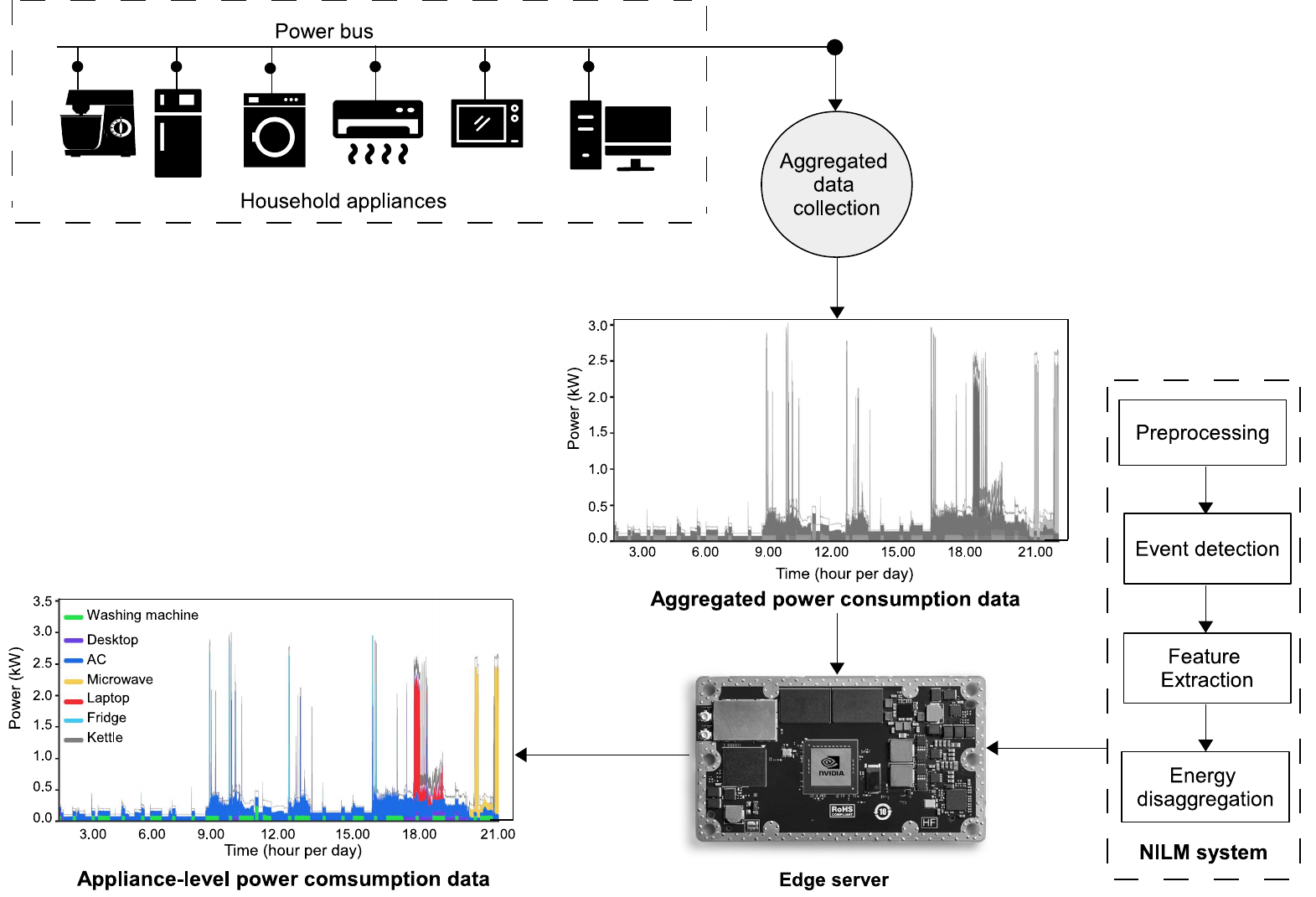}
\caption{Example of an edge NILM system implemented in the (EM)$^{3}$ framework using a Jetson TX1 server \cite{himeur2020robust}.}
\label{edge-NILM} 
\end{figure}

In \cite{himeur2020robust}, a robust NILM technique is proposed based on a multi-scale wavelet packet tree (MSWPT) feature extraction that is powered by an ensemble bagging tree (EBT) classifier. It has been implemented on an NVIDIA Jetson TX1 edge server \footnote{\url{https://developer.nvidia.com/embedded/jetson-tx1}}. Similarly in \cite{himeur2020effective1}, the authors introduce an efficient NILM method based on the aggregation of a (i) multi-decriptor fusion strategy, (ii) dimensionality reduction scheme using QR-decomposition, and (iii) decision bagging tree (DBT) classification model. Fig. \ref{edge-NILM} portrays the MSWPT-EBT based edge NILM implemented using the Jetson TX1 server.

\subsubsection{Edge-based non-technical loss (NLT) detection (A4)}
NTL, also known as energy fraud detection, pertains to unbilled energy consumption due to theft activities like smart meter bypassing or tampering \cite{glauner2016challenge}. As smart grids expand globally, security issues, especially NTL fraud, become paramount. While many NTL detectors exist, none can identify big data fraud within the grid. The study in \cite{han2019edge} introduces ENFD, a detection system powered by edge computing and big data analytics tools. Tests show ENFD can detect big data NTL frauds more efficiently than existing methods, operating six to seven times faster than the fastest known detector.
An edge data center, positioned between the data source and the central data center, can reduce data transmission and processing time. Identifying energy theft is vital for these centers to ensure reliable decision-making and reduce economic losses \cite{zhang2020energy}. The study in \cite{zhang2020energy} proposes a threshold-based detector for energy theft in edge data centers. The system uses a VAE-GAN-based feature extractor, K-means clustering for typical load profiles, and sets a threshold for detecting abnormalities. The research confirms that a converged VAE-GAN can understand real data's temporal and numerical relationships.

\begin{figure}[!t]
\centering
\includegraphics[width=1\columnwidth]{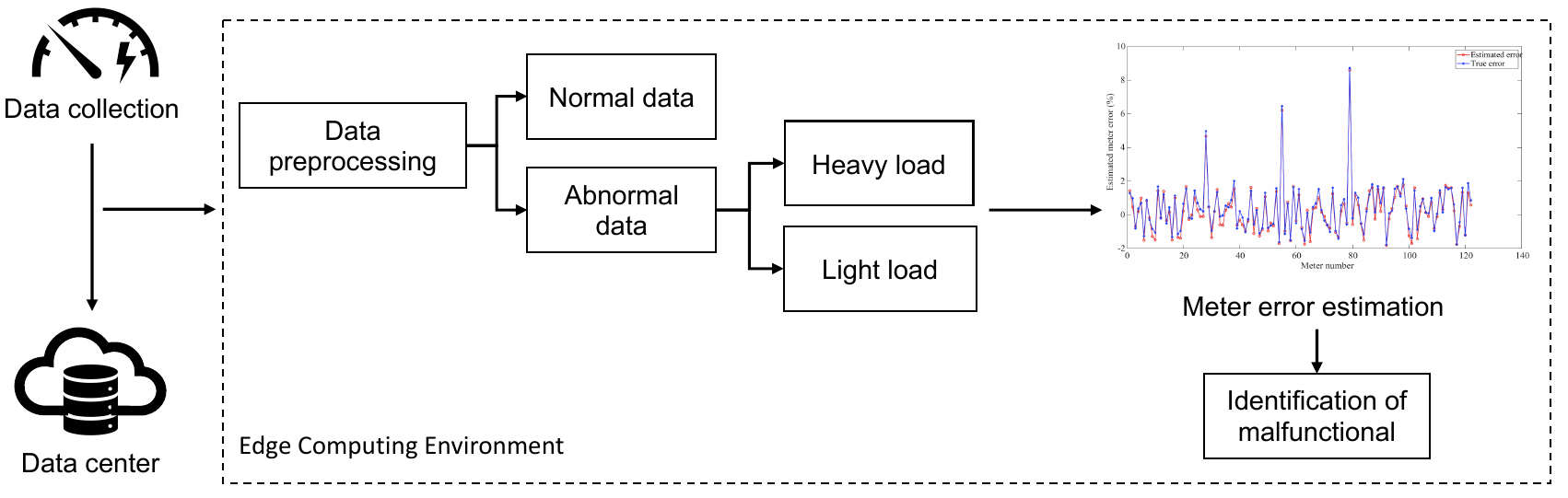}
\caption{Example of edge-based meter malfunction detection workflow \cite{liu2020remote}.}
\label{edge-malfunction-detection} 
\end{figure}

Energy fraud causes significant economic losses for utilities. In \cite{olivares2020machine}, different ML models, including Decision Tree Regressor (DTR), Logistic Regression (LR), Shallow Neural Network (SNN) and Multi-Layer Perceptron Regressor (MLPR) are introduced to detect irregular energy usage patterns indicative of fraud. These models, leveraging regression and classification over edge-fog computing, show promise for integration into smart metering systems.
Many global smart meters face the challenge of identifying high-volume failures. Manual inspections are costly and inefficient for widespread meters. The study \cite{liu2020remote} introduces an online method for detecting malfunctioning meters using their data. The system uses a decision tree to identify irregular data, then clusters it based on consumption patterns. A meter data matrix is then formed to determine meter errors. A recursive algorithm estimates meter errors, identifying those exceeding a set threshold as malfunctioning. The proposed method proved highly accurate in tests. A typical example of an edge-based meter malfunction detection is shown in Fig. \ref{edge-malfunction-detection}.

\subsubsection{Edge-based energy RSs (A5)} 
The vast growth of online information has led to the use of RSs to filter out unimportant details and suggest valuable items to users. In recent IoE-based energy efficiency systems, RSs have become integral \cite{sardianos2020real}. Predominantly, these RSs adopt the cloud-to-edge approach, processing data on cloud servers before sending results to edge devices like smartphones. However, issues like network bandwidth and latency can cause system delays, impacting user experience \cite{chen2019improved}. To better meet user expectations in real-time, processing directly on the edge can be more effective. \cite{gong2020edgerec} presents a pioneering edge-based RS, EdgeRec, designed for real-time user understanding and recommendations on edge mobile devices. The system also integrates a context-aware re-ranking for behavior attention networks to cater to diverse user preferences. Testing on the Taobao home page showed EdgeRec's efficacy in both offline evaluations and online outcomes.

Cloud-based recommendation systems (RSs) struggle to discern user needs effectively. MEC shifts computing resources from distant cloud servers to network edge servers, facilitating enhanced and personalized services. In \cite{sun2020convergence}, the integration of RSs with edge computing is extensively reviewed, shedding light on potential advancements. A unique perspective is offered in \cite{kotsogiannis2017directed}, where the RS recommends items that one user might gift or suggest to another. By blending individual user preferences with gift transaction data, the proposed algorithm accounts for the inclinations of both giver and receiver, leading to more tailored recommendations. This approach stands out when juxtaposed with typical group recommendation systems and edge tagging in social networks. Meanwhile, \cite{su2019edge} introduces a big data framework for promoting traditional cultural heritage. Beyond standard operations like querying and searching, this system proposes a novel user-centric technique to recommend cultural artifacts from varied repositories.

\begin{figure}[!t]
\centering
\includegraphics[width=0.6\columnwidth]{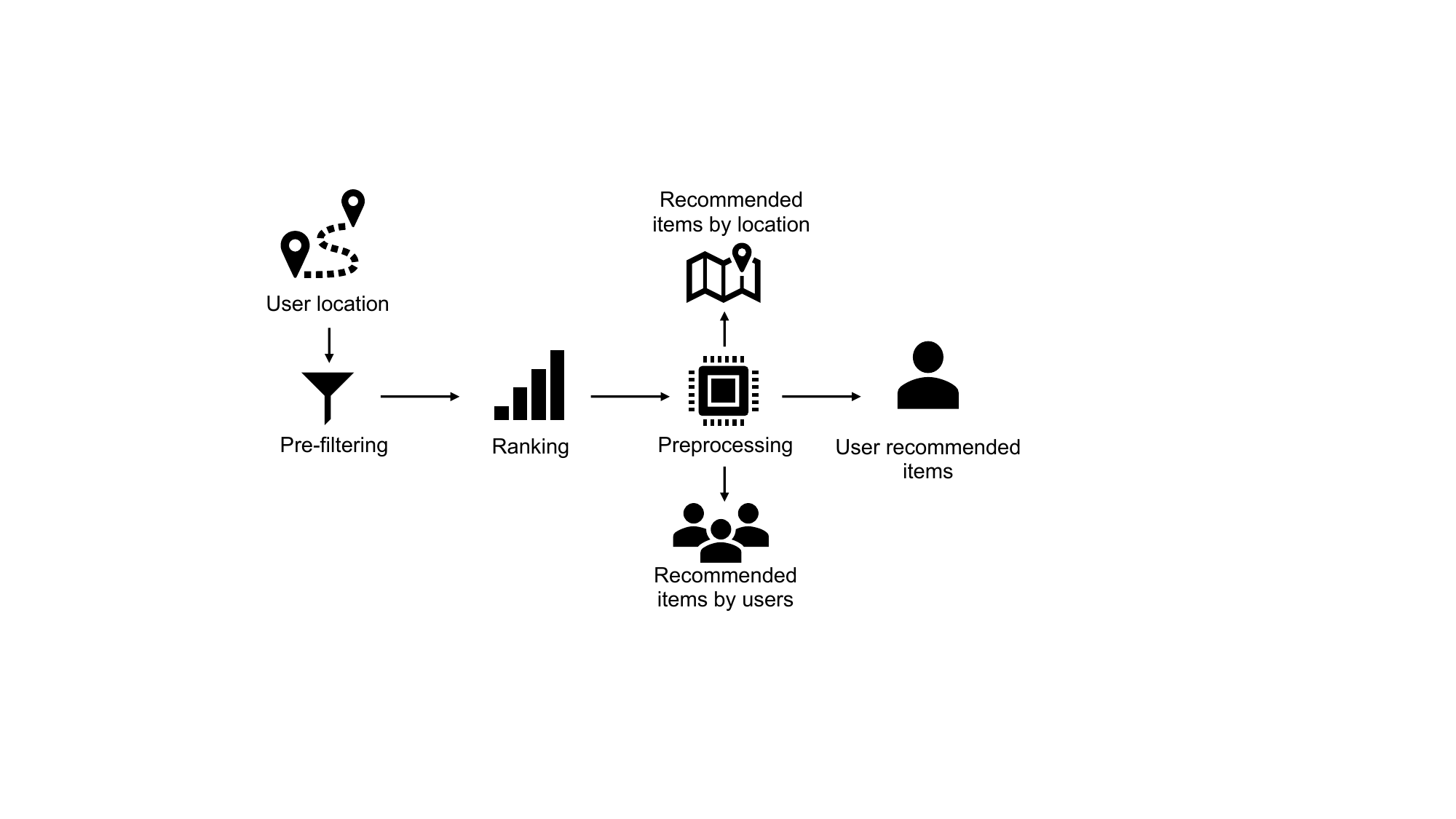}
\caption{Instance of a edge museums RS workflow \cite{su2019edge}.}
\label{edge-recommender-system} 
\end{figure}

Beyond the previously discussed IoE applications, edge-based RSs are being used in diverse scenarios. Specifically, \cite{su2019edge} utilizes a big data architecture to introduce a unique user-centric recommendation method for cultural artifacts. Despite the dominance of cloud computing, edge intelligence is harnessed through a smartphone app named "smart search museum". This app performs semantic searches and uses machine learning to suggest museums and other attractions to users based on their location, harnessing both joint recommendation strategies and edge AI capabilities. The effectiveness of the introduced system is affirmed by empirical results showcasing its accuracy and user satisfaction. 
\cite{alsalemi2021smart} introduces the Edge-IoE$^3$, an efficient and cost-effective platform designed for improving energy efficiency in homes. This platform, resembling a smart plug, comprises two parts: a unit for data collection and another for data processing. The collection unit uses various sensors to gather information on energy usage and environmental conditions like temperature, humidity, and room occupancy. The platform utilizes innovative micro-moment analysis through ensemble bagging trees to pinpoint specific energy consumption moments. These moments are then used to offer recommendations for better energy habits. 
Similarly, \cite{varlamis2022smart} introduces a mart-plug-based online RS that promotes energy-saving behaviors. By integrating sensor data, user habits, and feedback, it provides timely energy efficiency suggestions. User responses both activate energy-saving measures and refine future recommendations. Using a stacked-LSTM for multi-sensor data, the system achieved 93\% to 97\% accuracy in determining the best time for recommendations. Moving forward, \cite{sardianos2021emergence}  introduces a context-aware recommendation system for energy efficiency that offers personalized and explainable suggestions tailored to user habits. The recommendations emphasize either economic savings or ecological benefits and provide reasons for suggested energy-saving actions. A Telegram bot study showed a 19\% boost in recommendation acceptance when both types of persuasion were used, proving the potential of intelligent recommendations in promoting energy-saving behaviors.
The RS workflow is visualized in Fig. \ref{edge-recommender-system}.

\subsubsection{Edge-based energy prediction (A6)}
\begin{figure}[!t]
\centering
\includegraphics[width=0.37\columnwidth]{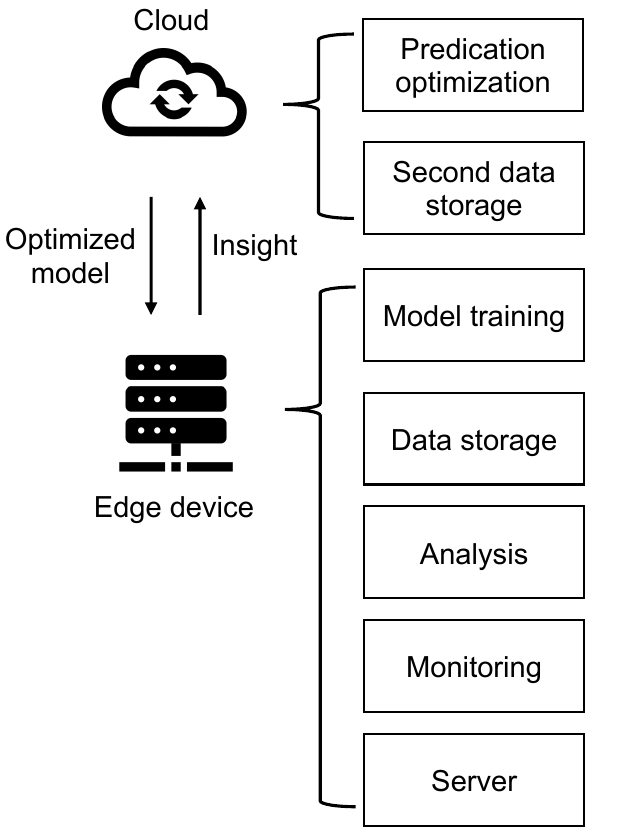}
\caption{Example of a workflow for edge-based energy predication \cite{lee2019energy}.}
\label{edge-prediction-system} 
\end{figure}

Energy management increasingly relies on real-time data interpretation for decision-making support, ensuring efficient energy production \cite{luo2019short}. Traditional cloud services can delay this crucial energy prediction. However, edge computing provides a more immediate solution. In \cite{lee2019energy}, a method using deep learning, specifically the LSTM network, is introduced for accurate energy consumption forecasting at the edge gateway, tested in an office environment. This approach demonstrates high precision in daily energy predictions.
In \cite{luo2019short}, an edge AI-based short-term energy prediction system is described. It divides data tasks between sensing nodes, routing nodes, and central servers. The system uses semantic processing and a tailored online DNN model for accurate IoE data analysis. A real-world building energy system application demonstrated the system's accuracy and reliability in real-time energy prediction. Fig. \ref{edge-prediction-system} presents an example of a workflow for edge-based energy predication \cite{lee2019energy}

In \cite{mocnej2018impact}, the impact of edge computing on IoT system energy consumption is explored. Through a case study, the authors assess the energy overhead of edge computation and suggest its application can extend battery life in certain devices. Abdellatif et al. \cite{abdellatif2020edge} introduce a multiaccess edge computing (MEC) architecture for energy-efficient and reliable remote health monitoring in response to the increasing patient numbers. Meanwhile, \cite{casado2020edge} presents an edge AI approach for green scheduling in smart buildings using an ESP8266 microcontroller.

% \cite{feng2020integrated}

\subsubsection{Edge-based occupancy detection (A7)}
\begin{figure}[!t]
\centering
\includegraphics[width=0.65\columnwidth]{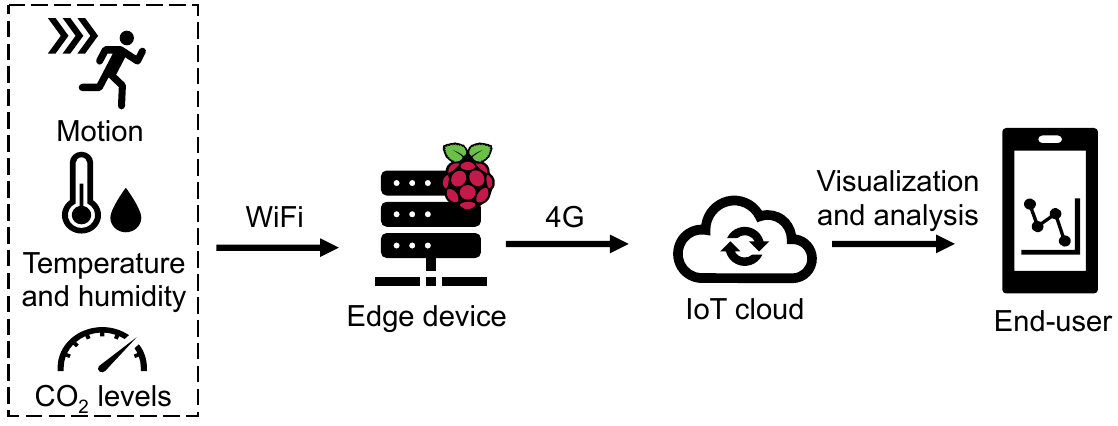}
\caption{On the edge occupancy detection method \cite{rastogi2020iot}.}
\label{edge-occupancy-detection} 
\end{figure}

Occupancy estimation is vital for optimizing demand-driven applications like smart lighting and heating, improving energy efficiency \cite{ke2020smart}. Workplace occupancy control can enhance environmental, visual, and air quality, resulting in significant savings \cite{sardianos2020model}. In \cite{zemouri2018edge}, the authors  use inexpensive temperature and humidity sensors to determine office occupancy and related energy consumption, exploiting the changes in environment due to human presence. Raspberry Pi platforms with these sensors were used and results were cross-checked against camera-derived occupancy data. The combined data can predict occupancy with an accuracy of up to 87\%.
\cite{tse2020deepclass} introduces an edge-based method for counting people in smart campus classrooms using cameras and Raspberry Pi platforms. Enhanced with image processing techniques, this method can be applied in various indoor settings without training. \cite{rastogi2020iot} presents an indoor occupancy prediction model using CO$_{2}$ and humidity data. Due to the large volume of IoT sensor data, latency issues arise during cloud transfers, which is addressed by running the model on edge computing. Fig. \ref{edge-occupancy-detection} illustrates the process of detecting occupancy on the edge.

\color{black}
\subsubsection{Power Generation}
IoE plays a pivotal role in energy generation, especially with the increasing integration of renewable energy sources \cite{bui2012internet}. Real-time data communication and advanced analytics provided by IoE can optimize the generation from sources like wind, solar, and hydro. 
The authors in \cite{iwendi2022combined} introduce a Power Generation and Electricity Storage Device (PGESD) for advanced microgrid systems. It focuses on intelligent home load control, leveraging deep learning and fuzzy logic for improved computation and reduced complexity. The system is adaptable to scenarios involving fluctuating power supply and microgeneration devices. It achieves a high precision and accuracy rate of 89\% and 92\%, respectively, offering a significant contribution to the development of Smart Buildings and Microgrids.
Moving on, \cite{wu2022integrated} addresses the growing demand for electricity amid energy scarcity by focusing on solar power technology. It emphasizes the importance of solar grid-connected inverters in optimizing power generation. The research tackles integration, efficiency, reliability, power consumption, and monitoring issues. It introduces a shadow radiant energy model, IoT-based networked platform, and ZigBee wireless sensor network to enhance distributed solar energy devices and promote the joint design of solar devices and buildings, benefiting the photovoltaic construction industry.
In the same direction, \cite{chen2019wind} addresses the increasing importance of wind turbine maintenance due to the rapid growth of wind power capacity. It offers insights into wind turbine fault diagnosis and prediction. The research evaluates fault diagnosis methods, including vibration analysis, electrical signal analysis, and pattern recognition algorithms. It also introduces a fault prediction approach combining physical failure models and data-driven models. Using deep learning within an IoT framework, the study successfully predicts and diagnoses wind power generation faults, demonstrating its practicality.

\subsubsection{Energy Distribution}
As energy moves from its generation point to the end consumers, the IoE helps in monitoring and controlling the flow. The ability to offload data and processes can be crucial here. For instance, if a particular section of the grid faces excess load, real-time data analytics, facilitated by offloading, can redirect energy flows, preventing outages.
For instance, Wu et al. \cite{wu2022computing} introduce a low-power method for Power Internet of Things edge devices. Utilizing computing offloading, it optimizes computing and communication resource distribution under energy constraints, addressing local resource limitations. The approach doubles equipment working time in zero-charge states. Integrating hibernation technology in the future can further prolong equipment operation and decrease power usage.
\cite{zhou2022joint} delves into the combined optimization of computing offloading and service caching in edge computing-based smart grids. By modeling the issue as an Mixed-Integer Non-Linear Program (MINLP) to diminish system task cost, the paper introduces the Collaborative Computing Offloading and Resource Allocation Method (CCORAM). Simulations reveal CCORAM's near-optimal performance and superiority over benchmarks. Future work will explore ES collaboration.
Moving on, \cite{li2020communication} introduces an edge computing-assisted smart grid fault detection system employing a lightweight neural network device near the equipment's edge for real-time monitoring. Addressing bandwidth overload and delayed feedback issues inherent in cloud-based approaches, the system optimizes communication and computation resource allocation. Simulations demonstrate enhanced data transmission speeds, reduced delays, and improved real-time performance compared to existing systems.

On another hand, AI at the edge can analyze consumption patterns and detect irregularities, potentially indicating energy theft. This helps utilities identify and address illegal energy consumption \cite{emadaleslami2023two}.
%\subsubsection{Energy Consumption}
%At the consumption end, IoE empowers consumers to be active participants in the energy market. Modern appliances can communicate with the grid, adjusting their operation based on energy prices and availability. Here, security becomes vital again; consumer appliances and smart meters must be safeguarded against hacking attempts.

\subsubsection{Renewable Energy Adoption}
Edge computing and AI play crucial roles in enhancing the efficiency, reliability, and cost-effectiveness of renewable energy systems \cite{li2018enabling}. They enable real-time monitoring and control, leveraging analytics-powered AI to optimize energy production and consumption based on demand patterns \cite{perin2021towards}. Additionally, AI-driven predictive maintenance minimizes downtime and enhances overall system efficiency. Energy storage benefits from optimization, predicting high-demand periods, and storing excess energy during low-demand times \cite{chen2021distributed}. Smart grids, empowered by AI and edge computing, facilitate seamless integration of renewable energy sources, automatically balancing supply and demand. This comprehensive approach reduces energy waste, and optimizes generation and storage, ultimately making renewable energy more economically viable and encouraging its broader adoption \cite{ku2020state}.

\begin{figure}[!t]
\centering
\includegraphics[width=0.95\columnwidth]{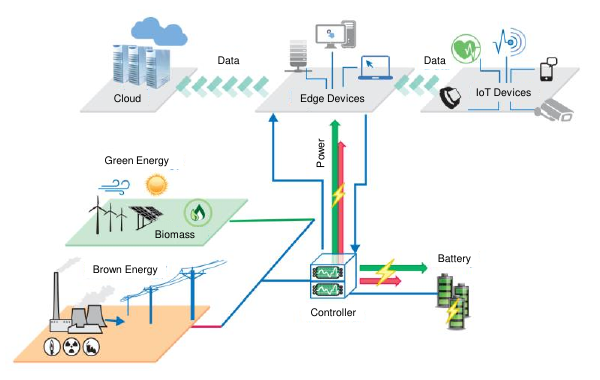}
\caption{\textcolor{black}{Flowchart of the framework proposed in \cite{li2018enabling} to integrate Edge AI into a microgrid.}}
\label{Edge_AI_RE} 
\end{figure}

For instance, Li et al. \cite{li2018enabling} propose a framework to facilitate collaboration between the energy supply system and edge computing, maximizing renewable energy utilization while enhancing service quality for time-sensitive IoT applications. An experimental system, combining a microgrid (solar-wind hybrid energy) and edge computing devices, has been introduced to validate the concept. Experiment outcomes revealed that renewable energy effectively powered the prototype system, ensuring the reliable operation of edge computing devices for the majority (94.8 percent) of the test duration.
Fig. \ref{Edge_AI_RE} portrays the flowchart of the framework used to integrate Edge AI into a microgrid.
Moving forward, \cite{perin2021towards} addresses energy efficiency in edge computing, focusing on reducing grid reliance and optimizing computing load distribution. Co-located with mobile network base stations, edge servers are powered by renewables, processing user workload with deadlines. The proposed predictive, online, and distributed algorithm achieves rapid convergence, optimizing energy use and potentially reducing renewable energy sales to the grid by up to 50\%.

\subsection{IoE security}
%The security aspect is paramount for IoE and related applications. For instance, ensuring that the communication lines transmitting data about generation metrics are secure prevents malicious attacks that could destabilize the grid \cite{kabalci2019smart}.
Security is paramount in the context of the Internet of Energy (IoE) as it safeguards critical energy infrastructure and data. With interconnected energy grids, smart meters, and renewable energy sources, vulnerabilities can have far-reaching consequences, including grid disruptions and privacy breaches. Robust security measures are essential to protect against cyberattacks, ensure data integrity, and maintain the reliability of energy systems. Moreover, security fosters trust among stakeholders, facilitating the widespread adoption of IoE technologies crucial for enhancing energy efficiency and sustainability \cite{kabalci2019smart}.
Numerous studies have addressed the security issues in IoE. For instance, \cite{song2014fpga} presents an FPGA-based Support Vector Machine (SVM) classifier for rapid data classification. Harnessing FPGA's advanced parallel computation capabilities, the system efficiently operates in both linear and non-linear modes based on classification dimensions. Simulated results highlight its efficacy, suggesting potential utility in enhancing Smart Grid communication security.
Similarly, Zhong et al. \cite{zhong2021data} address the challenges of cloud storage methods for power distribution IoT. It proposes a Data Security Storage method, leveraging a collaborative cloud-edge architecture for real-time processing. The Kademlia-based distributed data storage method ensures data security through homomorphic encryption and secret sharing. A security model for edge nodes based on noncooperative differential games is also introduced, optimizing defense strategies. Experimental results demonstrate superior query performance, resistance to network attacks, reduced storage/query delays, and enhanced data security.

\subsection{Smart-meter-based IoE}
\begin{figure}[!t]
%\centering
\includegraphics[width=1\columnwidth]{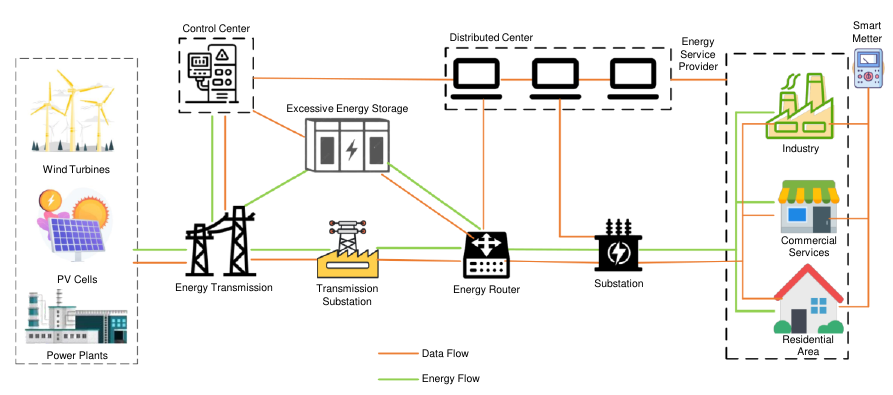}
\caption{\textcolor{black}{Smart metering infrastructure used in edge-AI-based IoE. }}
\label{SM-Infra} 
\end{figure}

Smart meters are pivotal in the IoE by enabling enhanced grid management, consumer empowerment, and the integration of various energy sources. They bridge the gap between traditional energy infrastructure and the modern digital landscape, ensuring the grid is more responsive, resilient, and efficient. Incorporating edge AI into smart meters transforms them from mere data collection devices to intelligent decision-making entities, driving the evolution of IoE and creating a smarter, more responsive energy network. 
Arenas et al \cite{arenas2020methodology} present an FPGA-based smart energy meter with a novel power quality analyzer in line with the IEEE 1459–2010 Standard, capable of remote monitoring under various electric power system conditions. The algorithms, coded in hardware description languages, are adaptable and defined by the grid's fundamental frequency. Experimental results showcase its versatility in different scenarios, validating real-time and parallel processing capabilities. Fig. \ref{SM-Infra} presents a typical architecture of the smart metering infrastructure used in edge-AI-based IoE.

\color{black}

\section{Techniques for fast inferences} \label{sec5}
\textcolor{black}{This section transitions to discussing methods that facilitate fast inferences in Edge AI systems, building upon the response to RQ3.} Typically, To allow the aforementioned applications to satisfy their latency demands, various architectures have been introduced to perform AI inference at a low cost. In this context, frameworks concentrated on using three major architectures are discussed, including (i) on-device computation, in which AI models are implemented on the end devices; (ii) edge server-based architectures, in which energy footprints and ambient data collected on the IoE network are transmitted to one or more edge servers for computation; and (iii) multi-modal computing, where the computation is executed on the end devices, edge servers, and the cloud simultaneously.

\subsection{On-device computation}
\begin{figure}[!t]
%\centering
\includegraphics[width=1\columnwidth]{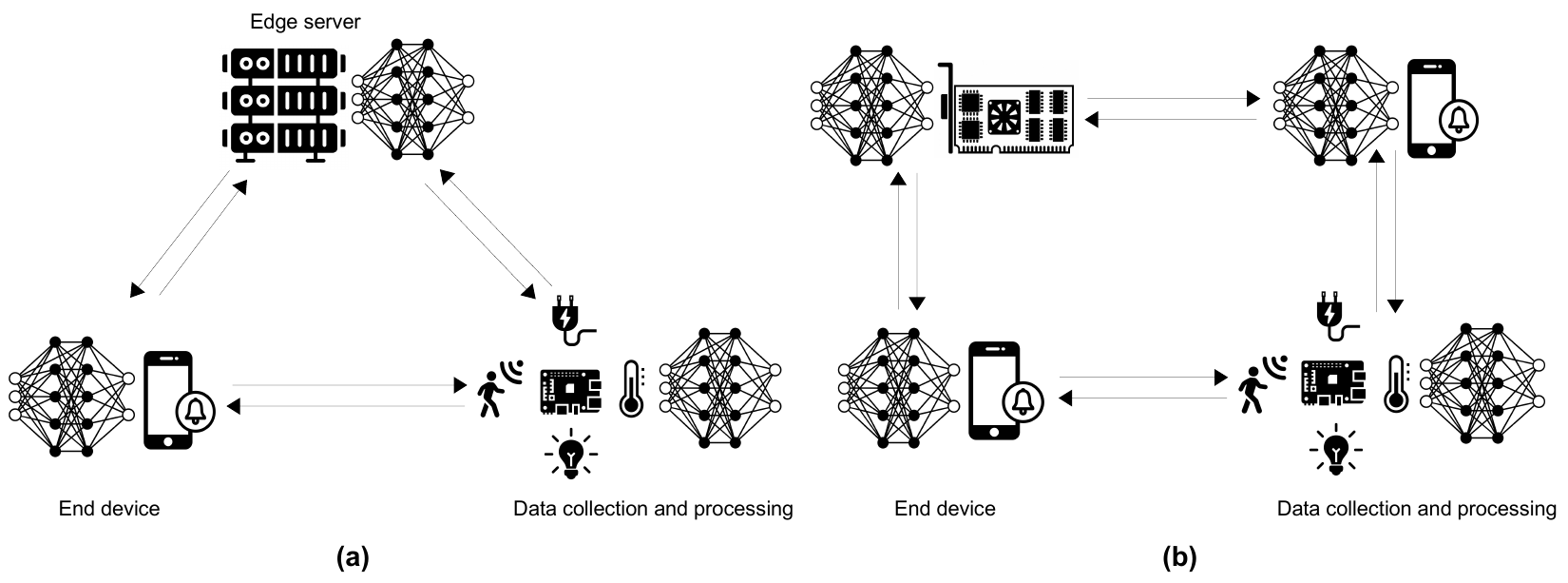}
\caption{Architectures for AI-based training on the edge for IoE applications: (a) centralized training, and (b) decentralized training}
\label{arch-AI-training} 
\end{figure}

A great research effort has been devoted to establish powerful mechanism for reducing 
the latency of AI algorithms when they are carried out on a resource-constrained device, as explained in Fig. \ref{arch-AI-training}. This helps in achieving some advantages over the edge ecosystem, such as latency reduction of AI algorithms while running on the end devices or edge servers. Therefore, we discuss some examples of effective hardware and AI models design:

\subsubsection{Model design} 
When developing complex AI models to be executed on resource-constrained devices, the AI community generally focuses on the design of appropriate algorithms having reduced number of parameters. This leads to reduce memory and running latency along with maintaining a high accuracy \cite{mocanu2016deep}. Accordingly, different techniques have been proposed to achieve this goal, especially when DL models are considered \cite{garcia2019estimation}. For example, among the popular DNN models designed for resource-constrained devices, we can find MobileNets \cite{howard2017mobilenets}, SqueezeNet \cite{iandola2016squeezenet}, solid-state drive (SSD) \cite{liu2016ssd} and YOLO \cite{redmon2017yolo9000}.

\subsubsection{Model compression}
It is another technique for enabling AI models on an edge device, especially DNNs \cite{cheng2018model,cheng2017survey}. Specifically, it relies on compressing existing DNNs while ensuring a minimum accuracy loss in comparison with original ones. Different model compression techniques exist, such as knowledge distillation \cite{gou2020knowledge,walawalkar2020online}, parameter pruning \cite{singh2019play,luo2017entropy} and parameter quantization \cite{singh2020leveraging,choudhary2020comprehensive}. An example of a use case scenario in IoE is presented in \cite{ahmed2020edge}, where an edge-based NILM framework is designed based on a phone-based implementation using MobileNet compressed by Tensorflow Lite. Accordingly, by compressing the model size, a smaller storage size is required on the end-user's device, and hence the use of RAM is reduced and inference time is lowered. Furthermore, the power consumption of the end-user's device is reduced because of reducing latency due to inference time reduction. This was achieved by using TensorFlow Lite that encompasses a post-training quantization option. The latter is performed thanks to the optimizattion that allows to lower the precision of the learning parameters, such as the MobileNet weights,
from their training-time 32-bit floating-point representations into much smaller and efficient 8-bit integer ones.

\subsubsection{Hardware}
Aiming at speeding up the inference of AI models, hardware producers try to leverage existing hardware such as CPUs and GPUs, in addition to manufacture custom ASICs for machine learning tools, specially those based on DL (Google’s tensor processing unit \footnote{\url{https://cloud.google.com/edge-tpu/}}). 
Other custom ASICs have been recently introduced, such as ShiDianNao, which concentrates on effective memory access
for reducing power consumption and latency. Typically, although it pertains to the DianNao group of DL accelerators, it has been directed into embedded devices, which is more practical in the context of edge AI computation. Furthermore, DNN acceleratos for FPGAs have also received a great interest due to the capability of FPGAs in providing efficient computation resources and keeping re-configurability \cite{adel2018accelerating}. Moving on, it is worth noting that conventional CPUs and GPUs are less power efficient compared to custom ASICs
and FPGAs, as the former have been developed for flexibly managing different workloads at the cost of
high power consumption.

\subsection{Edge Server Computation} \label{sec4-B}
Hardware speedup and compression methods could enable running AI models on edge devices, however, developing real-time IoE based applications that require large and powerful AI models is a challenging issue due to the existing of limited resources in terms of the power, memory and computation. Therefore, it is logical to search other alternatives, such as considering offloading AI computations from edge devices to other entities that have greater computing power, e.g. the edge servers or cloud \cite{carvalho2020computation,miao2020intelligent,guo2019toward, zhang2021efficient}.
The latter is not appropriate for real-time IoE applications requiring very low response times \cite{moghaddam2018fog}. On the flip side, edge servers are very convenient because they are close to end-users and could respond swiftly to their requests. Tho that end, this option is becoming the option one to developing real-time IoE applications \cite{lin2020survey,hong2019multi}.

\subsection{Computing across edge devices}
Even though edge servers could accelerate AI processing, the execution of AI models on them is not always needed as the intelligent offloading is a more practical alternative in some situations \cite{yu2017survey,wei2022delay,zhu2021dynamic}. Four offloading options are identified (i) binary offloading, which aims to decide whether to offload the entire AI model or not \cite{wei2022delay,zou2023privacy};  (ii) AI model partitioning, which refers to a partial offloading that aims at identifying what fraction of the AI model needs to be offloaded \cite{qiu2023dynamic}; (iii) combination of edge devices and cloud, which refers to hierarchical architectures that enabling offloading on combined edge devices, edge servers, and cloud \cite{yuan2023elect}; and (iv) distributed computation, in which AI computation is distributed across various peer devices \cite{utkarsh2016consensus,syed2020performance}.

\vskip2mm

\subsubsection{Binary offloading} 
The simplest approach to achieve that is via offloading the overall computation from end devices to the edge or cloud servers \cite{aujla2018mensus,chen2018dynamic}. Accordingly, end devices need to transmit their data to a close edge/cloud server and then receive the corresponding outputs following the server processing \cite{wei2022delay,zou2023privacy}. 

For instance, in \cite{liu2019intelligent}, a deep reinforcement learning (DRL) based energy-scheduling system for demand-side response is proposed with regard to the limitations of edge computing processing capacity. In this line, two kinds of DRL techniques have been deployed: an edge-DRL scheme and a cooperative DRL approach. The former relies on offloading the energy scheduling to an edge server before this one uses the DRL algorithm to identify optimal scheduling performance for devices. In the latter, the edge server offloads the DNN training to a cloud server to reduce the computational cost before adopting a deep Q-learning procedure using the computed Q-value from the cloud server. 

\vskip2mm

\subsubsection{AI model partitioning} 
Also termed as partitioning and offloading, this concept pertains to a fractional offloading scheme. It capitalizes on the distinct structure of an AI model, such as the layers in the context of DNNs \cite{zhou2019adaptive}. Taking DNNs as an example, model partitioning techniques allow for computation of certain layers on the device, while others might be computed by the edge/cloud server \cite{merenda2020edge,mohammadi2018deep}. In this framework, these techniques might reduce latency by utilizing the computing cycles of other edge devices. However, it is worth noting that communication latency due to intermediate data at the DNN partition point might offset the total net benefits \cite{teerapittayanon2017distributed,de2018partitioning}. A primary advantage of model partitioning is that after processing the initial layers of a DL model, the size of intermediate results shrinks compared to raw data, accelerating their transmission across the network \cite{elgamal2020serdab,ko2018edge}. This concept is termed as DNN model partitioning \cite{zhou2019distributing,li2018edge}.

Furthermore, various frameworks have emerged to partition convolutional neural networks (CNN) models with the objective of amplifying inference rates. For instance, \cite{dey2018partitioning} introduces a depth-wise input partitioning approach for CNNs, addressing the limitations of row/column or grid-based methods. Similarly, \cite{martins2019partitioning} proposes a CNN partitioning approach tailored for constrained IoT devices, significantly enhancing the inference rate and reducing communication operations between edge devices. Additionally, \cite{zhao2018deepthings} presents DeepThings, a solution for deploying adaptive distributed CNN-based inference methods on IoT devices with restricted computing resources. Specifically, DeepThings employs a scalable fused tile partitioning (FTP) approach to reduce memory footprint while ensuring parallel computation. On the other hand, \cite{de2018partitioning} suggests a Kernighan-and-Lin (KLP) based partitioning method for CNNs, enabling efficient distributed execution across multiple IoT edge devices. This method achieves 4.5 times less communication compared to other methods, such as TensorFlow. Lastly, \cite{kim2020energy} showcases a hardware management method to divide neural networks into sets of consecutive layers and pipeline their execution using various processing units. This strategy, dubbed NeuroPipe, allows the embedded processor to dispatch faster inferences at reduced voltage and frequency, enhancing energy efficiency while maintaining high computational performance.

\subsubsection{Combination of edge devices and cloud} 
Although computation offloading to the cloud can limit the real-time IoE applications, a wise use of effective cloud computation resources can help in decreasing the overall processing cost. Techniques of this category frequently rely on AI partition, in which some tasks will be executed in the edge server, cloud and/or end device. For example, in \cite{li2018learning}, the authors divide a DL model into two parts and execute them on (i) edge servers and the cloud. The former has been used for computing the first layers of the model, while the latter has been utilized for computing the higher layers. Specifically, when the edge servers received the input data, they performed a lower layer DL processing, before sending the intermediate results to the cloud. Following, the latter sent back the final results to the end devices after
computing the higher layers. In this context, the cloud has been responsible on (i) executing the computational demanding tasks, (ii) increasing the edge server’s request processing rate, and (iii) decreasing network traffic between the cloud dn edge servers. Moving forward, in \cite{teerapittayanon2017distributed}, the computation of a DNN model is has been distributed over multiple platforms, i.e. end devices, edge servers and the cloud. Moreover, a model compressing approach and a fast exiting idea have been deployed to avoid that computation requests always reach the could.

\subsubsection{Distributed Computation}
While the aforementioned techniques focus on offloading the compute tasks from end devices to more powerful platforms, i.e. cloud or edge servers, another category of edge AI computing schemes consider the problem as a distributed computing perspective, in which the AI computation is distributed across
different helper edge devices \cite{ben2019demystifying}. In this context, various works have been proposed, such as the Scalable distributed dl training in \cite{wang2019scalable} and MoDNN \cite{mao2017modnn}, which both are based on distributing DL executions using in-depth partition on lightweight end devices, e.g. Android smartphones \cite{zhao2018deepthings}, Raspberry Pis \cite{mao2017modnn}, Jetson Nano and Jetson TX1 \cite{fang2020cachenet}.

\subsection{Private inference}
When data are transmitted from edge devices (in the edge network) to edge servers, as explained previously (Section \ref{sec4-B}), sensitive data are vehiculated, such as, occupancy patterns, ambient conditions and energy footprints foorprints, which can be used to infer the behavior of the end-users, their habits and their presence/absence times. Therefore, this lead to security and privacy issues since there is a risk to leak or hack these data. Even though edge AI logically enhances the privacy and strengthens the security through limiting data transmissions to cloudlet platforms via the Internet, there are still some issues between edge devices and edge servers.The latter can be further be improved using:

\subsubsection{Add Noise to Data}
Numerous studies have explored methods to safeguard sensitive data transmitted from edge devices to edge servers for inference by introducing noise. In this context, a differential privacy scheme is introduced in \cite{wang2018not}, which adds noise, specifically following the Laplace distribution, to the output of a DL model. Similarly, the work in \cite{mireshghallah2019shredder} presents the Shredder method, a privacy preservation technique that learns to add noise distributions. This method aims to significantly reduce the transmitted data's information content while retaining high inference accuracy. Furthermore, Wang et al. in \cite{wang2019privstream} introduce PrivStream, a privacy-preserving strategy tailored for IoT-based big data analytics frameworks that employ edge computing. This approach is bifurcated into two segments, one deployed on an IoT device and the other on an edge server. The former obfuscates sensitive data by infusing it with Laplace noise, while the latter reconstructs the obscured information from the noisy data stream, ensuring that it can be used for inference without compromising privacy.

\subsubsection{Secure computation}

Cryptographic methods provide a robust solution to ensure private inference. The primary goal of securing the computation of AI models lies in ensuring that edge devices obtain inference results without gaining any knowledge about the AI models \cite{zhang2018data,azzouzi2023novel}. Simultaneously, the edge server processes the data without gaining insights into the sensitive information. To put it another way, both edge devices and edge servers work to compute an AI prediction, 
$p(X,Y)$, where $X$ denotes input samples (e.g., energy consumption observations) known only to edge devices, and $Y$ represents the AI model parameters exclusively available to edge servers. Secure computation, therefore, facilitates the calculation of $p(X,Y)$ without either party having to access the other's data.
Homomorphic encryption is one strategy to achieve secure computation. With this method, data is encrypted before being transmitted from edge devices to edge servers. Computations then take place on this encrypted data, as presented in \cite{gilad2016cryptonets}. This reference introduces a technique that converts trained neural networks into "CryptoNets." These are adapted neural networks designed to work with encrypted data. CryptoNets allow the transmission of data between edge devices and servers (or even cloud servers) in an encrypted form. This encryption ensures data confidentiality since the edge servers (or cloud servers) lack the decryption keys required to access the original data.

\section{AI Training On The Edge} \label{sec6}
\textcolor{black}{This section explores the training aspects of AI on the edge, which is a critical component. It discusses the frequency of training updates,  size of training updates, decentralized communication protocols, and private training. It also provides answers to RQ3.}

\subsection{Frequency of Training Updates}
Minimizing the cost of IoE applications is crucial to facilitate widespread adoption. A practical approach to achieving this objective involves reducing the communication expenses of edge devices. This can be accomplished by decreasing both the frequency and size of communications. We first delve into distributed training techniques that aim to cut down the frequency and timing of communications. Subsequently, we assess methods that focus on shrinking the size of the training data.
Two principal schemes have emerged to synchronize updates with the central edge server. The first, synchronous stochastic gradient descent (SSGD), entails updating the parameters of edge devices in unison after they compute gradients on their respective training data batches \cite{shi2018dag}. In contrast, asynchronous stochastic gradient descent (ASGD) allows each device to independently relay its parameter updates to the primary server. Both SSGD and ASGD come with their respective advantages and drawbacks \cite{du2021asynchronous}. While SSGD often converges to superior solutions, its real-world performance can be hampered due to its inherent waiting period for slower devices in every iteration. On the other hand, ASGD typically exhibits a quicker convergence compared to SSGD. However, this swiftness can sometimes be a double-edged sword; if parameter updates are based on outdated information from edge devices, it might lead to convergence on suboptimal solutions.

\subsection{Size of Training Updates}
As previously highlighted, the size of training updates plays a significant role in bandwidth consumption, alongside the frequency of these updates. The bandwidth demands can be considerable, especially when noting that model sizes often span hundreds of megabytes and require multiple rounds of communication. This demand becomes particularly salient in edge computing applications, where last-mile bandwidth, such as in wireless and access networks, may be notably restricted. In this context, we delve into gradient compression schemes designed to minimize the size of the training updates sent to the primary server. The literature identifies three primary gradient compression approaches, which we discuss as follows:

\subsubsection{Sparse Compression}
Sparse compression refers to the technique where gradients are selected based on their highest $k$ absolute values before transmission to the main server \cite{wangni2017gradient}. For instance, \cite{strom2015scalable} presents an approach that sends only the gradients exceeding a specific threshold to the main server. In practical applications, selecting an appropriate threshold is challenging, as it can vary widely across different models and even between layers within a single model \cite{stich2018sparsified}. Similarly, \cite{aji2017sparse} employs a fixed compression rate instead of a predefined threshold to determine which gradients to communicate. Here, a subset of gradients with the highest magnitude is sent, while the remaining gradients accumulate in a residual. Notably, using a low compression rate minimally impacts the final model's accuracy and convergence speed. To address the stale information issue and increase sparsity in the initial communication rounds, \cite{lin2017deep} introduces a curriculum approach tailored to the model. The deep gradient compression (DGC) technique is developed, achieving compression rates between 1/600 and 1/270 across various training models without compromising final accuracy or convergence speed. This technique has subsequently been applied to training various RNN and CNN models with large-scale datasets.

\subsubsection{Quantitative Compression}
Quantitative compression pertains to the local quantization of gradients, often using randomization, to reduce the bit count required for representation \cite{karimireddy2019error}. In \cite{wen2017terngrad}, the authors present TernGrad, a method that stochastically quantizes gradients into ternary values. This method achieves a compression rate of 1/16, but its accuracy significantly deteriorates when trained with large datasets. The convergence of this approach is underpinned by assuming bounded gradients. In another work, \cite{alistarh2016qsgd} delves into finding the optimal balance between accuracy and gradient precision. The research identifies theoretical bounds on compression rates attainable through dense quantization and subsequently implements a quantization-based SGD scheme to ensure convergence. Meanwhile, \cite{bernstein2018signsgd} introduces signSGD, a distributed training method where each computing edge unit quantizes gradients into binary representations. The main server then aggregates these gradients through a majority vote. It is noteworthy that quantitative compression methods, at their best, can reach a maximal compression rate of 1/32, as detailed in \cite{jiang2018linear}. Similarly, \cite{seide20141} proposes a 1-bit quantization approach, empirically showcasing the feasibility of quantizing gradients to a single bit without compromising convergence speed, especially when quantization errors are aggregated.

\subsubsection{Communication Delay}
This strategy capitalizes on the idea that edge devices can perform several local model updates before communicating the results, which effectively reduces communication frequency \cite{chen2018lag}. In line with this, \cite{mcmahan2017communication} introduces a federated averaging scheme where each computing node undertakes multiple iterations of SGD to compute the gradients, rather than communicating after every single iteration. It has been demonstrated that using this method, the convergence speed remains largely unaffected despite delaying communication. More notably, there is a reduction in the required communication rounds by factors ranging from 10 to 100 across various RNN and CNN algorithms. Progressing further, \cite{konevcny2016federated} merges the communication delay with both probabilistic quantitative compression and random sparse compression. Here, computing nodes are either constrained to learn random sparse gradients or enforce random sparsity upon them. Although this approach has demonstrated commendable compression performance across different CNN and LSTM models, it does result in a noticeable decrease in both the final accuracy and convergence speed.

\subsection{Decentralized Communication Protocols}
Within decentralized networks, edge intelligence can optimize computational distribution, especially in mesh networks, where smaller devices collaboratively handle computational tasks \cite{hijawi2020lightweight}. This setup bolsters security through immediate data encryption. A study by \cite{hijawi2020lightweight} unveils a framework with lightweight key-policy attribute-based encryption (KPABE) for wireless mesh networks, emphasizing its scalability, adaptability, and enhanced security for IoT nodes. This framework's application in a school displayed its prowess in energy conservation optimization, integrating IoT components. On another note, \cite{wang2019social} presents "social edge intelligence (SEI)," merging AI and human intelligence to address significant scientific challenges in advanced computing. Inspired by AI's rapid growth in mobile technologies and crowdsourcing's potential, the SEI approach amalgamates insights from multiple disciplines to address evolving SEI application challenges.

\subsection{Private Training}
Considering the SSGD approaches in the context of security and privacy implications associated with gradient communication, such methods can be effective even when the training information from one edge device is shared with others. While training typically enhances privacy in real-world applications—primarily by preventing edge devices from sharing directly collected data—there remains a significant risk of private information leakage during gradient exchanges between devices \cite{ryffel2018generic,aono2017privacy}. Hence, there is a pressing need for advancements that bolster privacy preservation. In this section, we delve into two primary categories of privacy preservation mechanisms. The first involves adding noise to the training data updates, while the second centers on utilizing secure computation techniques for training.

\subsubsection{Adding Noise to Data Training Updates} 
Edge AI models may inadvertently retain traces of training data, which could encompass sensitive information. Such residues could potentially allow adversaries to extract the said information through meticulous analysis of the model \cite{abadi2016deep}. A countermeasure to this is introducing noise to the training updates. For instance, \cite{du2018differential} introduces a differential privacy mechanism where the sanctity of training data is upheld by infusing Laplace perturbations to mask sensitive details. Subsequently, two algorithms—objective perturbation and output perturbation—are conceptualized to facilitate this differential privacy. Similarly, \cite{ma2020safeguarding} implements noise addition to the training data on the client side as a precaution against revealing sensitive information. \cite{mao2018learning} amalgamates differential privacy with model partitioning to obfuscate training updates, thereby enhancing privacy. Here, the preliminary layers of a DNN model are processed on an edge device, intertwined with perturbations, and then relayed to the edge server. Lastly, \cite{zhang2018privacy} explores the effects of introducing diverse perturbations to input data prior to the training process \cite{zhang2020detecting,liang2020super}.

\subsubsection{Secure Computation} 
Secure computation facilitates the development of protective protocols designed for training and inferring intricate edge AI algorithms. This is particularly relevant when amalgamating data from multiple sources, which can often conflict with data privacy concerns \cite{merenda2020edge,wagh2019securenn}. In this context, Agrawal et al. propose a secure computation methodology in \cite{agrawal2019quotient} for a dual-phase training and evaluation of DL models, named QUOTIENT. This approach securely trains convolutional and residual layers, forming crucial components of contemporary DNNs. In a related vein, \cite{wagh2018securenn} introduces three-party and four-party secure computation protocols tailored for various neural network components, such as rectified linear units, matrix multiplication, normalization, MaxPool, among others. These innovations contribute to the creation of three-party and four-party information-theoretically secure protocols that are suitable for training and predicting using DNNs, CNNs, and other deep learning frameworks.

To summarize what have been discussed, Table \ref{Summary-edgeAI} outlines the relevant edge AI solutions for IoE applications. These works are described with reference to the implemented AI technique, used edge platform, advantages, targeted application and metrics used to evaluate the performance.

\begin{center}
\begin{longtable}{
    m{0.5cm}
    m{2.8cm}
    m{3.2cm}
    m{4cm}
    m{0.8cm}
    m{2.6cm}
}
\caption{Summary of the reported work on ChatGPT describing area of the study, applications, objectives and key findings of the research.}
\label{Summary-edgeAI}\\
\hline 
Ref. & Implemented technique & Edge platform & Advantages &  Appl. & Evaluation metric \\ \hline
\endfirsthead
\hline
\multicolumn{5}{c}{{Table \thetable\ (Continue)}} \\
\hline
Ref. & Implemented technique & Edge platform & Advantages &  Appl. & Evaluation metric \\ \hline
\endhead
\hline
\endfoot

\small

\cite{sirojan2019embedded} & PCA & edge node (NI sbRIO-9637) & enable
real-time energy data analysis & A1 & n/a \\ 

\cite{himeur2020emergence} & DNN + micro-moments  & edge server (ESP8266
NodeMCU V1.0) & low-cost real-time anomaly detection & A2 & accuracy=93.86\% \\ 

\cite{schneible2017anomaly} & DNN + autoencoders & mobile edge devices & 
perform the computation in parallel on the edge devices & A2 & F1=99.05\% \\ 

\cite{mohamudally2018building} & ARIMA & edge router & support real-time
anomaly detection & A2 & n/a \\ 

\cite{marchioni2020subspace} & improved PCA & edge server & implementation
on low resource devices & A2 & running time = 49.7ms \\ 

\cite{xu2019data} & OCSVM + statistics and data-driven scheme & edge network
& high detection accuracy and effective computational performance & A2 & PR-AUC=76.83\% (KDD dataset) \newline PR-AUC=79.83\% (UNSW-NB15 dataset) \\

\cite{zhu2019mobile} & multi-task LSTM neural network & mobile edge devices
& multi-dimensional anomaly detection  & A2 & accuracy=90\% \\

\cite{shah2016edgecentric} & compression-based approach & edge-attributed
networks &  & A2 & PR = 87\% \\

\cite{hussain2019mobile} & DNN & mobile edge devices & improve network's QoS
and user's QoE & A2 & accuracy = 98.8\% \newline precision = 99.07\% \\ 

\cite{luo2018arrays} & nearest-neighbor search based LSH & mobile edge
devices & computational savings & A2 & 60-300$\times$ faster than existing approaches with competing accuracy \\ 

\cite{ngo2020adaptive} & DNN + single-step Markov decision & hierarchical
edge computing & reduces anomaly detection delay by 84\% & A2 & accuracy = 98.9\% \newline F1 = 83.3\%\\ 

\cite{lin2019edge} & RNN & edge server (NVDIA Jetson TX2) & accelerate the
training time almost 120  times faster than the traditional model& A2 & 100\% of true
alarm rate  \\

\cite{ezeme2019deep} & DNN (LSTM) & edge devices & improve the security on
the edge nodes & A2 & regression error (n/a) \\

\cite{ahmed2020edge} & DNN (MobileNet + TensorFlow Lite) & mobile edge
device (emulator) & cheaper and scalable energy disaggregation solution on android device  & A3 & 
average relative error RE$_avg$ = 8.46\%\\

\cite{cao2019achieving} & FHMM + noise addition & hybrid edge-cloud  & 
ensure privacy-preservation & A3 & F1 = = 72\% \\ 

\cite{hernandez2020design} & CNN & edge-computing node (SoC architecture) & 
reduce further bandwidth requirements & A3 & n/a\\

\cite{xiang2019iot} & disaggregation using extended Kalman filter & edge
server (smart plug) & cost-effective solution  & A3 & n/a \\ 

\cite{liu2020secure} & rule-based event detection & edge gateway & safely
monitor the overall load  & A3 & accuracy = 98.89\% \newline F1 = 96.04\% \\ 

\cite{han2019edge} & recursive least square & edge devices & six times
fastest thatn existing frameworks & A4 & n/a \\

\cite{zhang2020energy} & VAE-GAN & edge data center & computationally
efficient and robust & A4 & precision = 97\% \\ 

\cite{liu2020remote} & K-means clustering + RMSE & edge devices (smart
meters) & High fraud detection accuracy & A4 & RMSE = 0.21\% \\

\cite{olivares2020machine} & DTR, LR, SNN, MLPR &  Raspberry Pi Model 3B+ & executed properly in embedded devices with limited computing capabilities & A4 & MAPE = 11.99\% \\

\cite{su2019edge} & knowledge-based collaborative filtering & mobile edge
devices & distributed computing on Hadoop and NoSQL repositories & A5 & accuracy = 97.7\%\\

\cite{luo2019short} & data-driven, online DNN model & edge perceptual nodes
& reduce the computational pressure on the central server& A6 & precision > 92\% \\

\cite{mocnej2018impact} & linear regression & edge devices (NodeMCU ESP-12E)
& suitable for the constrained devices with the limited battery & A6 & n/a \\

\cite{zemouri2018edge}  & EWMA & edge server (Raspberry Pi) & low-cost,
non-intrusive occupancy detection & A7 & accuracy = 87\%\\ 

\cite{tse2020deepclass}  & DNN & edge server (Raspberry Pi) & suitable for
the constrained edge devices with the limited battery & A7 & accuracy \\

\hline
\end{longtable}
\end{center}

\color{black}
\subsection{Federated Edge AI}
Federated learning has emerged as a promising approach in the context of the IoE to enable collaborative and privacy-preserving ML across a network of decentralized edge devices. Typically, typically it is widely use in in edge-AI-based IoE for several compelling reasons: (i) in IoE, data privacy is paramount. Energy consumption patterns, user behavior, and grid-related data are often sensitive and subject to regulations. Federated learning allows edge devices to collaborate on model training without sharing raw data, thus preserving user privacy and complying with data protection laws \cite{yoo2022fuzzy}; (ii) IoE encompasses a wide range of devices, including smart meters, sensors, and distributed energy resources. These devices generate data at the edge, and centralizing this data for traditional machine learning can be impractical and inefficient. Federated learning allows training to occur locally on these devices, harnessing their computational capabilities \cite{venkataramanan2022forecast}; (iii) energy management and grid optimization in IoE require low-latency responses and real-time decision-making. Federated learning's distributed nature aligns with the edge's capacity for quick data processing, enabling timely responses to changing conditions in the energy grid \cite{liu2021federated}; (iv) IoE networks are often vast and growing, making scalability a crucial concern. Federated learning can scale seamlessly with the addition of new edge devices, as each device can participate in model training independently without the need for centralized infrastructure \cite{lin2021privacy}; (v) edge devices typically have limited computational and energy resources. Federated learning is resource-efficient as it offloads most of the model training work to the edge, reducing the need for heavy computational infrastructure and minimizing energy consumption; (vi) in IoE, network connectivity can be unreliable. Federated learning's decentralized approach ensures that even if some edge devices experience connectivity issues, the overall training process can continue. This redundancy enhances the robustness of IoE applications; (vii) federated learning allows for customized model updates on individual edge devices. This is valuable in IoE scenarios where energy consumption patterns can vary widely between users or devices. Personalized models can improve the accuracy of energy management and optimization; (viii) by keeping data localized and using secure aggregation techniques, federated learning enhances the security of IoE applications. It reduces the risk of data breaches associated with central data repositories; and (ix) Federated learning aligns well with data privacy regulations such as General Data Protection Regulation (GDPR). It enables organizations to demonstrate compliance by minimizing data exposure and enhancing user consent management.

In the context of IoE, FL is implemented as follwos. Initially, a global machine learning model is created and initialized with certain parameters. This global model serves as the starting point for training across edge devices. Edge devices, such as smart meters or energy management systems, perform local model training using their own data. This training is done while maintaining data privacy, as raw data remains on the edge device. After local training iterations, edge devices generate model updates based on the knowledge gained from their data. These updates typically include information about model weights and gradients. The model updates from multiple edge devices are securely aggregated. Secure aggregation techniques, such as homomorphic encryption or differential privacy, ensure that individual updates remain private while allowing the computation of an aggregated global update. The aggregated model update is applied to the global model. This process iterates over multiple rounds, allowing the global model to learn from the collective knowledge of all edge devices without centralizing the raw data.

Numerous studies have been proposed in the literature to discuss the use of Federated learning in edge IoE systems. For instance, the technique of non-intrusive load monitoring, which identifies appliance activity from energy use, can inadvertently expose user behaviors. While local differential privacy offers a more user-centric privacy solution than centralized methods, achieving a balance between privacy and utility remains challenging. AI-based obfuscation methods offer potential solutions, but their computational demands often exceed the capabilities of most power IoT devices.
In  \cite{cao2020ifed}, the IFed system was introduced, leveraging FL to allow resource-rich electricity suppliers to support Power IoT users. The system harmonizes local differential privacy, data utility, and resource use, and addresses the privacy concerns of machine learning model transfers between suppliers and consumers. This ensures heightened privacy, especially for sensitive users, demonstrating that IFed meets Power Internet users' privacy requirements.
Computational offloading decisions in IoT systems are complicated by factors like federation, resource management, and varying radio conditions. Ren et al. \cite{ren2019federated} propose using multiple DRL agents on different edge nodes to aid these decisions. To enhance feasibility and reduce transmission costs, FL trains these agents in a decentralized manner. Experimental results underscore the effectiveness of this combined DRL and FL approach in advanced IoT setups.

%\section{Discussion of current key challenges and future perspectives}

\color{black}
\section{Limitations and Current Challenges} \label{sec7}
\textcolor{black}{This section provides a critical analysis of the Edge AI landscape and highlights current challenges. It touches upon security, computation, processing, and dependability issues. It also answers RQ4.}

Edge AI, though a burgeoning concept, faces a myriad of challenges. Despite having addressed certain issues more effectively than other technologies and offering distinctive contributions to enhancing the quality of service (QoS), there remain significant concerns, particularly in the domains of data storage, computation, network activity, usage, and localization.

\subsection{Security and Privacy}
It is worth noting that the security and privacy measures commonplace in cloud computing cannot be seamlessly transplanted into the realm of edge AI, primarily due to its geo-distribution, inherent heterogeneity, and reliance on wireless connections \cite{himeur2022latest}. In the context of IoE applications employing edge AI, the diverse data types sensed via IoT devices are processed across distinct edge devices or distributed edge servers. Such a setup inherently poses higher security and privacy risks than traditional cloud servers \cite{sayed2021intelligent}. Moreover, these edge servers, potentially under varying operators, might need to collaborate to accomplish mutual tasks, amplifying the security and privacy challenges \cite{alwarafy2020survey}. Furthermore, there are scenarios where collaboration between edge servers is crucial for undertaking a variety of tasks, especially when operated by different entities, thereby exacerbating the existing concerns. Collectively, the primary reasons behind the heightened security and privacy challenges in edge AI encompass:

\begin{itemize}
\item Given that intelligence and computational tasks are executed on edge nodes proximate to end-users, these nodes handle a significant volume of privacy-sensitive data \cite{ding2022roadmap}. Consequently, if these nodes are compromised or if there is a data breach, the ramifications can be grave \cite{kamruzzaman2022new}.
\item Edge nodes, by their nature, possess limited computational resources when contrasted with larger cloud platforms. This limitation hinders their ability to deploy sophisticated security measures \cite{lv2022edge}.
\item The dynamic nature of edge computing environments, stemming from the high mobility of edge nodes and end-users, creates a landscape that's ever-evolving \cite{mahmud2022ifogsim2}. This fluidity offers opportunities for malicious entities to infiltrate the network. Furthermore, establishing robust security protocols becomes intricate when multiple domains intersect, encompassing device providers, data generators, end-users, and more \cite{zhang2022deep,zhang2023effective}.
\end{itemize}

\begin{figure}[!t]
\centering
\includegraphics[width=0.65\columnwidth]{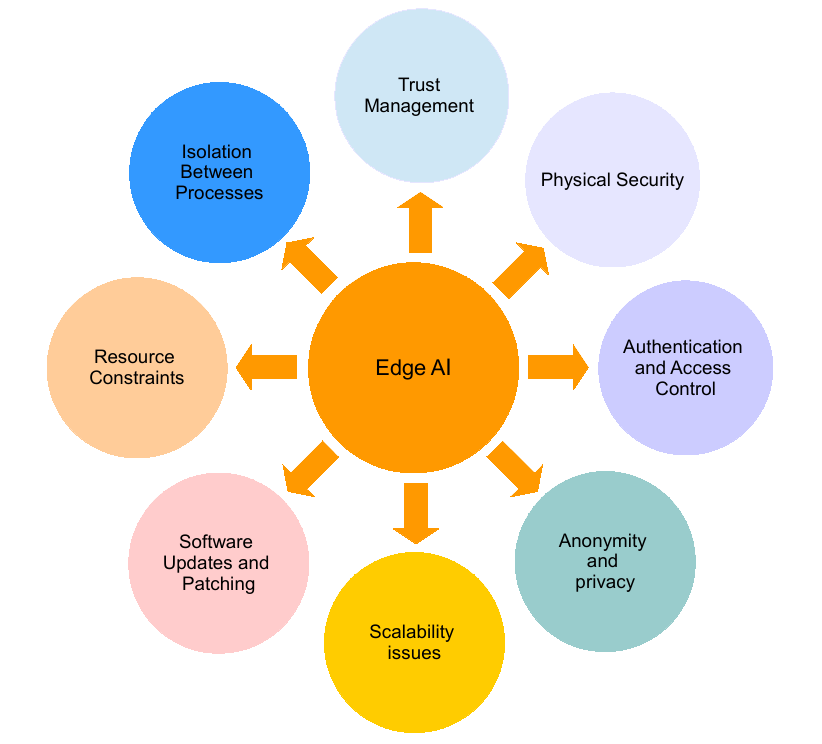}
\caption{\textcolor{black}{A summary of the main security and privacy issues in edge AI systems.}}
\label{privacy-issues} 
\end{figure}

Furthermore, while differential privacy offers an appealing framework for balancing privacy conservation and data utility, its deployment in the IoE context for shielding sensitive inferred data poses two main challenges. The initial challenge arises from the reality that integrating noise into IoE data streams can significantly degrade the utility of the data. The subsequent challenge stems from the inherent difficulty in efficiently sampling IoE data streams, largely because the distribution of such streams is typically unpredictable.
To encapsulate the salient security and privacy hurdles inherent in edge AI applications, we reference Fig. \ref{privacy-issues}. Typically, these challenges can be summarized as follows: (i) Trust Management: It refers to the process of assessing and managing trustworthiness among various nodes or entities in a system. It involves defining policies to build, evaluate, and revoke trust and the mechanisms to ensure that entities in the network adhere to these policies \cite{kim2019security}; (ii) Physical Security: Edge devices, being widespread and sometimes located in easily accessible areas, are vulnerable to physical tampering or theft \cite{kumar2023digital}; (iii) Authentication and Access Control: With a decentralized approach, ensuring that only authorized devices and users can access the system becomes complex. Secure authentication and access control mechanisms are crucial \cite{suciu2021sealedgrid}; (iv) Anonymity and Data Privacy: Unlike centralized cloud systems where data can be uniformly protected under heavy security protocols, edge devices can be more exposed. If not encrypted or handled correctly, sensitive information can be intercepted or altered \cite{singh2023bsems}; (v) Scalability issues: Given the sheer number of devices in some edge networks, security solutions need to be scalable. Implementing security on each device might be resource-intensive or infeasible \cite{khubrani2023blockchain}; (vi) Software Updates and Patching: Keeping edge devices updated with the latest security patches can be challenging, given their distributed nature. An unpatched device can become a potential entry point for attacks \cite{amjad2023performance}; (vii) Resource Constraints: Edge devices often have limited computational resources. Implementing heavy security solutions can affect their primary functionality \cite{sha2020survey}; (viii) Isolation Between Processes: Ensuring that different applications and processes running on an edge device are isolated from one another is important to prevent a compromise in one application from affecting others \cite{li2022edge}.

Addressing these hurdles, especially in terms of bolstering mechanisms that safeguard user and data privacy and ensuring secure communication between edge devices, is a focal point of ongoing research endeavors.

%Data Provenance: Ensuring the integrity and origin of the data is crucial. Attackers might inject false data, leading to wrong decision-making by the AI system.

\subsection{Computation}
Edge computing servers, in contrast to cloud-based systems, often can't support high-end AI and deep learning models due to hardware constraints \cite{xu2020edge}. If the industry focuses on creating scaled-down software versions, it may hinder the integration of extensive data analytics on edge AI platforms. The growing demand for more computational and storage capacities, along with the complexities in communication, becomes harder to meet with the increase in data volume \cite{yousefpour2019all}
For instance, in NILM applications, low-cost commercial smart meter devices cannot support advanced energy disaggregation techniques due to their limited computational capabilities. Additionally, these devices cannot record energy data at high-frequency resolutions. Given these limitations, there is an urgent need to address the shortcomings of current smart meter devices by developing more powerful and intelligent measurement nodes.

\subsection{Processing}
Designing edge AI functionalities to process and analyze diverse data collected by IoT devices using intelligent algorithms and to offer services to end-users is a challenging task. Such designs can significantly increase the volume of uploaded data, leading to heightened latency, especially when dealing with new or mixed types of data from various integrated devices \cite{deng2020edge}. Consequently, multiple AI model partitioning schemes have been explored; however, some of these can inadvertently increase the communication streams between the IoT devices.

\subsection{Dependability}
Edge AI embodies the idea of relocating intelligence and computing resources closer to the edge devices within the IoE and the originating data sources, akin to a CPU cache's function. While this can augment bandwidth and minimize latency for IoE applications, it often comes at the expense of dependability and capacity. A core reason behind this is that edge devices typically aren't as consistently maintained, powerful, reliable, or robust as centralized server-class cloud platforms \cite{bakhshi2019dependable,ramirez2020system}.
To be more specific, the dependability challenge encompasses hardware failures, such as crashes or hangs, software issues linked to performance, and violations of real-time constraints. While edge AI can enhance real-time execution by curtailing end-to-end latency, it is akin to a cache miss situation: overloaded or malfunctioning edge devices can push latency beyond acceptable thresholds \cite{bagchi2019dependability}.

\subsection{Interoperability and Standardization}
\begin{figure}[!t]
\centering
\includegraphics[width=0.95\columnwidth]{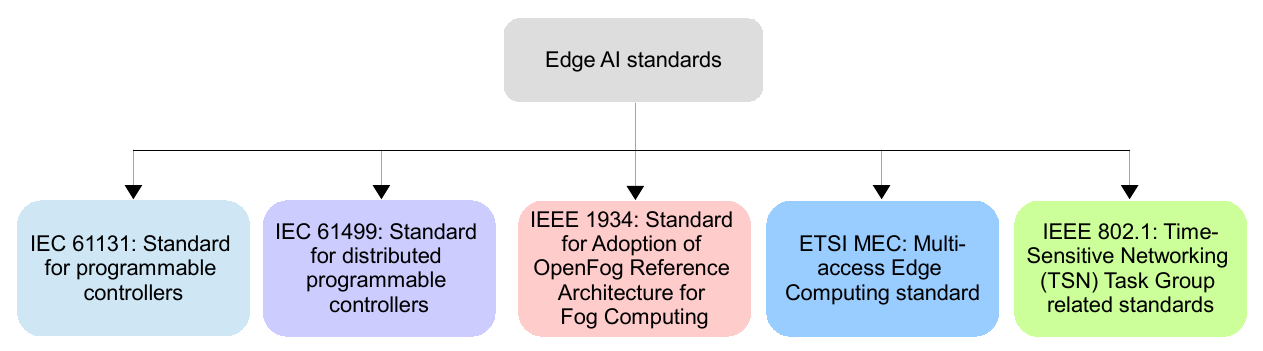}
\caption{A summary edge paradigms registered or are currently under consideration.}
\label{edge-standards} 
\end{figure}

The IoE ecosystem comprises various devices from different manufacturers. Ensuring that they all "speak the same language" and can work in tandem is crucial but can be challenging \cite{himeur2022recent}.
The development of edge AI frameworks spans a multifaceted domain. Beyond the myriad technical challenges, economic and political factors, such as reshaping the ICT sector or bolstering a nation's digital economy, also influence the evolution and application of edge AI solutions. Currently, edge AI grapples with inconsistent terminology, standards, norms, and interpretations \cite{hamm2019edge}. Although recent years have witnessed several initiatives aiming to address edge AI's complexities—including collaborations between research institutions and industries, community-driven software projects, and company-initiated software undertakings—a unified standard remains elusive. Consequently, potential adopters continue to seek edge AI solution providers they can rely on. Fig. \ref{edge-standards} depicts some of the prevalent edge AI standards, either established or under deliberation, spanning areas like edge computing, fog computing, and MEC.

\subsection{Data Consistency}
The advent of edge computing has brought forth a paradigm shift in how data processing and storage are approached. Instead of relying solely on centralized data centers, edge computing pushes these tasks closer to the source of the data, i.e., the edge devices \cite{keshari2022survey}. While this decentralized approach offers numerous advantages, including reduced latency and increased responsiveness, it also introduces challenges, with data consistency being a prominent one \cite{surianarayanan2023delineating}. This problem is due to the following reasons:
\begin{itemize}
\item Nature of Edge Devices: Edge devices, by their very nature, are diverse and dispersed. These can range from IoT sensors, smartphones, routers, to industrial machines. Each device might have its own local storage and processing capability, leading to multiple data versions, formats, or states across the network \cite{al2022ai}.
\item Concurrency Issues: If multiple edge devices try to update a piece of shared data concurrently, it can create conflicts. For example, two devices might report different readings for the same metric, leading to ambiguity about which reading is accurate \cite{xu2022certificateless}.
\item Network Partitions: Given that edge devices might be spread out geographically and connected over varied network types, there is a risk of network partitions. During such events, some devices might become isolated and continue to operate on outdated data. When the partition is resolved, merging the data changes from the isolated devices with the current data state can be challenging \cite{wang2022privacy}.
\item Latency Variances: Different edge devices might experience varied latency levels, causing asynchronous data updates. A device with higher latency might update data later than one with lower latency, potentially overwriting more recent data changes \cite{yu2022augmented}.
\item Data Durability and Availability: Edge devices might not always have the same data redundancy measures as centralized data centers. If an edge device fails or loses its data, restoring the lost data and ensuring its consistency with the rest of the system becomes a concern \cite{lu2023auction}.
\item Bandwidth Limitations: Constantly syncing data between edge devices and the central cloud requires bandwidth. In environments where bandwidth is limited or expensive, maintaining real-time data consistency can be a challenge \cite{zeng2022influences}.
\item Operational Complexities: Managing data consistency requires sophisticated algorithms and protocols. Implementing and maintaining these, especially in a decentralized and dynamic environment like edge computing, can be operationally complex \cite{sicari2022insights}.
\item Security and Compliance: Ensuring data consistency also means ensuring that all data, regardless of where it is stored or processed, adheres to the same security and compliance standards. This can be more challenging when dealing with numerous edge devices that might have different security postures \cite{mei2022blockchain,himeur2022blockchain}.
\end{itemize}

\subsection{Data Tempering}
Limitations related to data tampering in IoE systems are a critical concern that can have far-reaching consequences. Typically, the vulnerability of data integrity is critical in IoE systems. Intruders with malicious intent can manipulate transmitted data, including critical variables like pricing information \cite{khattak2020dynamic}. For instance, an attacker may alter peak-time energy usage prices to reflect the lowest rates. This can lead to unintended consequences such as overloading the power network due to increased demand during what appears to be a low-cost period \cite{kamal2018light}. This poses a substantial threat to the reliability and accuracy of the information within the system. Moreover, in the context of distributed state estimation, where multiple smart devices and sensors are involved in monitoring and controlling energy-related parameters, data tampering can severely impact the integrity of the estimation process. Erroneous data injection can occur, leading to incorrect state estimations. These inaccuracies can propagate through the system, potentially causing cascading failures.

Besides, the consequences of data tampering can be dire. In the worst-case scenario, if malicious data injections go undetected and unmitigated, they could lead to a widespread blackout. Imagine an attacker manipulating data to create a false sense of surplus energy, leading to the improper allocation of resources and eventually resulting in a grid failure affecting numerous consumers \cite{cintuglu2019secure}.

\color{black}

\section{Future Perspectives} \label{sec8}
\textcolor{black}{In this section, we shed light on current research directions and pinpoint areas that are drawing significant research and development attention for both the immediate and distant future. Specifically, this section answers RQ5.}

\subsection{Edge-to-Edge Cooperative AI}
The proliferation and advancement of IoE applications in domains like the smart grid, smart buildings, and smart cities have led to a soaring demand for AI-driven services. While existing frameworks primarily revolve around the collaboration between edge devices and cloud-edge AI, they often grapple with subpar AI learning efficiency \cite{zhang2019edge,hong2019multi,ren2019survey}. In this context, edge-to-edge cooperative AI emerges as a promising solution to this persisting challenge. However, many of the current research endeavors are primarily centered on offloading the computation of AI-related tasks, overlooking its intrinsic nature as a complex, brain-like process. Such an approach tends to subject raw data to intricate processing, resulting in increased latency and compromised learning quality \cite{jha2019iotsim}.

\subsection{5G networks}
The introduction of 5G communication networks marks a notable milestone for the deployment of edge AI-driven IoE applications. Notably, edge devices utilizing 5G boast considerably higher speeds and dramatically reduced latency when contrasted with cloudlet platforms. Furthermore, studies indicate that 5G networks can achieve up to 90\% greater energy efficiency per traffic unit compared to their 4G predecessors. As a result, edge AI operations become not only more efficient but also more energy-conservative than those on cloudlet platforms. Recognizing the potential, several utility and tech companies have initiated robust efforts to roll out 5G-integrated edge solutions. Such advancements enable effective management of big data across a network of edge devices \cite{abdelzaher2019panel}, thereby equipping edge AI-driven IoE solutions with the nimbleness required to rapidly deliver services to end-users.
Concurrently, this evolution has spurred the development of Mobile Edge Computing (MEC) within 5G, facilitating the integration of computing servers either within or in close proximity to the base station, bringing them closer to mobile subscribers.

\subsection{Deep reinforcement learning}
Leveraging the computational power at the edge of IoE networks, Edge AI can adeptly handle complex tasks in IoE using computation offloading. Capturing comprehensive IoE data, which covers consumption details, environmental conditions, and available resources, presents a challenge, especially due to the tight interdependence of IoE devices. Deep Policy Reinforcement (DPR) offers a way forward, particularly in contexts with limited IoE network information \cite{dai2019deep}. This approach not only addresses core challenges but also extends to energy cost and conservation tasks.
For instance, \cite{yu2019deep} addresses energy cost minimization in smart homes, bypassing the need for a thermal dynamics model, and instead harnesses DPR to navigate complexities such as model instability and parameter uncertainties. In a similar vein, Hua et al. \cite{hua2019optimal} employ a DPR-centric strategy for energy conservation in the IoE sector, reframing the energy conservation challenge as a constrained control problem, without relying on a fixed model for green energy outputs. Their innovative approach underscores the adaptability and effectiveness of DPR in addressing intricate issues in the IoE domain.

\subsection{Blockchain edge AI}
Giving that the end-users’ data is very sensitive and can be illegally accessed when edge-based IoE application are developed, the design and integration of security and privacy preservation mechanisms are becoming among the great research directions, in which scientists are investigating a significant effort. In this line, blockchain that is an emerging technology, has been used for the context of edge AI applications. This is because it can significantly enhance the security of edge AI based system for IoE applications since it enables the interaction of only trusted edge devices/nodes with each other \cite{xiong2018mobile}. Actually, particular importance is being put to develop security mechanisms based on permissioned blockchain for edge AI frameworks, such as \cite{gai2019permissioned,honar2021hyperledger,jayasinghe2019trustchain,pahl2018decision}.

Moreover, since edge AI systems have distributed architectures, decentralized security mechanisms that rely on hybrid SDN-blockchain is considered as a cutting-edge technology that can improve significantly the security and privacy preservation \cite{hu2020securing}. Accordingly, SDN technique is adopted for providing continuous monitoring of edge AI networks, whereas the blockchain strategy is embedded for guaranteeing a decentralized security and hence avoiding single point failures \cite{rathore2019blockseciotnet,yazdinejad2020energy,rahman2021smartblock}.

On another hand, and from the IoE perspective, blockchain has been used in smart grid to manage energy distribution in which the blockchain distributed consensus has been utilized ti verify demand response \cite{pop2018blockchain}. Moreover, the Ethereum coin has been deployed for paying for energy while Solidity has been employed to implement smart contracts. In a similar manner, the authors in \cite{ferrag2020deepcoin} introduce the DeepCoin, which refers to the combination of blockchain and learning-based schemes. The former refers to robust peer-to-peer energy system using the practical Byzantine fault tolerance algorithm, which prevents smart grid attacks by generating blocks using hash functions and short signatures. While the learning approach consists of an intrusion detection system based on recurrent neural networks to detect attacks and illegal transactions in the blockchain energy network.

\subsection{Software-Defined Networking for edge AI}
SDN is grabbing significant interest in recnet years for its smart capability of reconfiguring edge devices and routing traffic of edge AI based networks. Moreover, it provides a powerful and secure mechanisms of authentication and access control \cite{sharma2018softedgenet,li2020secured}. In additon, it can be used in some situations to fill the gap when aggregating edge computing and cloud computing. Specifically, it can be utilized for acting to decide on if some specific tasks might be uploaded and processed on the edge or on the cloud \cite{li2018adaptive,wang2019sdn}. On the other hand, SDN based distributed authentication technology for edge AI applications is also a promising research direction \cite{wang2019software,baktir2017can,muthanna2019secure}.
Moving on, designing SDN-based handover authentication management methods that ensure low computational delays and using less communication resources  is another research orientation that attract a growing attention \cite{wang2019sdn,duo2020sdn,chuang2020network}.

\subsection{Next Generation EVs}
The surge in the number of electric vehicles (EVs) on our roads each year is a testament to a global shift, with more people opting for them over conventional gasoline-powered vehicles, encompassing cars, motorcycles, and more. This not only underscores the growing environmental consciousness but also has implications for our power grids: EVs can act both as a power source and as a strain. Concurrently, the emergence of autonomous vehicles (AVs), or self-driving cars, represents a seismic shift in transportation. Unlike traditional vehicles, AVs function without human intervention, leveraging cutting-edge sensor technology to mitigate accidents, reduce energy consumption, diminish pollution, alleviate road congestion, and minimize human errors. These vehicles, equipped with sensors, radars, and cameras, constantly survey their surroundings—be it roads, obstacles, potential hazards, or traffic flow—ensuring a safer and more efficient drive. Importantly, they are eco-friendly and adept at navigating traffic congestions, determining obstructions, and charting the quickest route to a destination.

Furthermore, with their capability to supply power back to the grid during periods of high demand, the integration of EV power systems with the Internet of Things (IoT) is gaining traction. In tandem with this, the Internet of Energy (IoE) paradigm is pivotal for ensuring demand flexibility in an era dominated by EVs. Propelling the flexible energy transformation is vital given the escalating adoption rates of EVs. By 2023, light-duty EVs are projected to number around 6.4 million, as per Navigant Research. This symbiotic relationship between EVs and the power grid is not limited to vehicle-to-grid (V2G) interactions but also amplifies grid-to-vehicle (G2V) engagements, ensuring a two-way exchange that benefits both.

\subsection{LLMs for improving the adoption of edge AI in IoE}
The advent of LLMs and generative chatbots has ushered in a transformative era in computational linguistics and artificial intelligence. Marrying vast datasets with intricate algorithms, these models are capable of producing human-like text, paving the way for enhanced natural language processing tasks \cite{alqahtani2023emergent}. From facilitating nuanced conversations in customer service to aiding researchers in complex data synthesis, driving content creation in the media sector to assisting in educational platforms, LLMs and chatbots have found applications across diverse fields, revolutionizing the way we interact with, and harness, information \cite{farhat2023analyzing}.
LLMs, such as ChatGPT, represent a convergence of expansive data analysis and linguistic comprehension \cite{sohail2023future,sohail2023using}. Their capacity to process, understand, and generate human-like text from massive datasets makes them instrumental in various technological domains \cite{sohail2023decoding}. When applied to the rapidly evolving landscape of the Internet of Energy (IoE) combined with edge AI, they promise transformative advancements. LLMs like ChatGPT have the potential to significantly enhance the application and utilization of edge AI in the IoE \cite{shen2023large}. They adeptly synthesize vast amounts of research and data, enabling a comprehensive understanding of the latest in edge AI techniques and methodologies. These models can recommend optimizations for algorithms and facilitate predictive maintenance by analyzing data patterns. They play a role in scenario analysis, predicting outcomes from integrating edge AI into IoE scenarios. In residential settings, LLMs can elevate user experiences by providing real-time information and energy-saving tips. Their proficiency in data preprocessing aids in dataset annotation and cleaning. Furthermore, LLMs serve as valuable educational tools, assisting in training and offering insights during the prototyping and validation phases. Their knowledge base encompasses the latest security protocols, making them assets in recommending safety measures for IoE infrastructure. Moreover, they guide the integration of edge AI with other emerging technologies, paving the way for innovative IoE solutions.

\section{Conclusion} \label{sec9}
Developing and applying edge AI on IoE attempts to (i) reduce the network congestion by bringing the computation and intelligence to the end devices, (ii) support real-time implementations by reducing the latency due to data transmission to cloud servers; and (iii) reduce power consumption in comparison with wireless transmission of data to the gateways or cloudlet servers. To highlight the contributions made in this regard, this article overviewed the current state-of-the-art for edge AI frameworks for IoE applications. Accordingly, different aspects have been discussed, including the potential of edge AI on IoE, techniques for fast inferences, AI training on the edge. Accordingly, this review focused on describing the edge potential impacts on IoE, especially those related to the improvements in terms of the latency, real-time analytics, scalability, information security and privacy, automated decision making and reduced cost. In addition, techniques used for fast inferences are deeply described, such as on-device computation, edge server computation, computing across edge devices and private inference. Moving forward, a set of examples demonstrating the use of edge AI in IoE applications has been presented, where the ease of use and effectiveness of the proper edge AI platforms have been discussed.
Therefore, it has been clearly seen, that the current state of development of edge AI for IoE applications is advancing rapidly but its initial complexity requires more research work to thoroughly understand all possible benefits and overcome the main drawbacks. Typically, using edge computing will help in developing a variety of solutions able to comply at least partially with a plethora of requirements, among them privacy preservation, computational complexity, power consumption, scalability, etc. 

Nevertheless, like any emerging technology, Edge AI for IoE is not devoid of challenges. Issues related to security, processing capabilities, interoperability, and data consistency require meticulous attention to ensure seamless integration and operation.

Looking forward, the horizon is filled with promising prospects. Concepts such as Edge-to-Edge Cooperative AI and Federated Edge AI hint at the collaborative future of the technology, leveraging collective intelligence to achieve greater results. The introduction of 5G networks, combined with cutting-edge techniques like deep reinforcement learning and blockchain, promise to elevate Edge AI's capabilities even further. Furthermore, the potential of software-defined networking and next-generation EVs opens up new arenas for innovation. The role of LLMs and generative chatbots also stands out as an influential factor in promoting the adoption of Edge AI within the IoE ecosystem.

Lastly, it is worth noting that the merger of Edge AI and IoE is set to redefine our technological landscape. By addressing current challenges and harnessing upcoming innovations, we can anticipate a future where smart, decentralized, and highly efficient systems become an integral part of our everyday lives.

\section*{Acknowledgements}\label{acknowledgements}
This paper was made possible by National Priorities Research Program (NPRP) grant No. NPRP14S-0401-210122 from the Qatar National Research Fund (a member of Qatar Foundation). The statements made herein are solely the responsibility of the authors.

\end{document}